# STABILITY ANALYSIS OF ROTATING BEAMS RUBBING ON AN ELASTIC CIRCULAR STRUCTURE

N. Lesaffre*, J-J. Sinou, F. Thouverez

Laboratoire de Tribologie et Dynamique des Systèmes UMR CNRS 5513
Ecole Centrale de Lyon, 36 avenue Guy de Collongues, 69134 Ecully, France

## ABSTRACT

This paper presents the stability analysis of a system composed of rotating beams on a flexible, circular fixed ring, using the Routh-Hurwitz criterion. The model displayed has been fully developed within the rotating frame by use of an energy approach. The beams considered possess two degrees of freedom, a flexural motion as well as a traction/compression motion. In-plane deformations of the ring will be considered. Divergences and mode couplings have thus been underscored within the rotating frame and in order to simplify understanding of all these phenomena, the degrees of freedom of the beams will first be treated separately and then together. The dynamics of radial rotating loads on an elastic ring can create divergence instabilities as well as post-critical mode couplings. Moreover, the flexural motion of beam rubbing on the ring can also lead to mode couplings and to the locus veering phenomenon. The presence of rubbing seems to make the system unstable as soon as the rotational speed of the beams is greater than zero. Lastly, the influence of an angle between the beams and the normal to the ring's inner surface will be studied with respect to system stability, thus highlighting a shift frequency phenomenon.

Keywords: Stability analysis, rotating beams, circular elastic ring, divergence, mode couplings, rubbing.

## I. INTRODUCTION

Problems caused by loads moving over elastic structures occur rather frequently. The case of rotating structures has been given special attention. A flexible rotating disk excited by loads can experience instabilities, which for instance can occur when using a circular saw, as widely studied by Mote [1,2], with computer memory storage disks, studied by Iwan and Stahl [3], Iwan and Moeller [4] and Crandall [5] among others, or in the field of brake systems, Ouyang *et al.* [6], Chambrette and Jezequel [7]. In all these examples, the studied system was composed of a rotating disk whose in-plane vibrations were considered rubbed on its plane by stationary loads or by a fixed disk rubbed by rotating loads. Post-critical instabilities have thus been identified and their leading parameters determined. Few studies however could be found that focus on a rotating ring rubbed by loads on its inner surface. Canchi and Parker [8] recently investigated parametric instabilities due to rotating springs on a circular ring. This kind of system can be applied, for instance, in planetary gears or turbo machinery, in the case of contacts between the rotor blades and the casing [9]. This study will thus present a stability analysis of a flexible ring rubbed on its inner surface by rotating beams with two degrees of freedom: flexural motion and traction/compression. To better understand this phenomenon, the two degrees of freedom will first be separated and then studied together. In the first section, the ring will be excited by beams featuring only the traction/compression degree of freedom rotating on its inner surface. Afterwards, the beams considered will possess just a flexural degree of freedom. Finally, these two degrees of freedom will be studied in combination with one another. This paper will conclude with an examination of the influence of a constant angle between each beam and the normal to the inner ring surface.

## II. THE MODEL

The model considered in this study consists of a flexible ring rubbed by one or several beams on its inner surface, as depicted in Fig. 1a, in the case of one rotating beam. These Euler-Bernoulli beams have two degrees of freedom in the rotating frame, i.e. in the frame attached to the beams: traction/compression motion $\vec{u}_t$, and flexural motion $\vec{\upsilon}_f$. An energy method is used to develop the model; hence, the degrees of freedom of the $j^{th}$ beam are expressed by the following Ritz functions $u_{t_j}(x,t) = u_{t_j}(t)\sin\left(\frac{\pi x}{2R_{stat}}\right)$, corresponding to the exact traction/compression mode shape of a clamp-free beam, and $\upsilon_{f_j}(x,t) = \upsilon_{f_j}(t)\left(1-\cos\left(\frac{\pi x}{2R_{stat}}\right)\right)$ for its flexural degree of freedom, with $x$ being the local axis along the beam and $R_{stat}$ the ring radius. Concerning the ring, its in-plane flexural vibrations are considered, i.e. two degrees of freedom are considered in the rotating frame: radial displacement $\vec{u}_s(\phi,t)$, and tangential displacement $\vec{\omega}(\phi,t)$, with $\phi$ being the angular position of the mass centre of a ring's cross-section in the rotating frame. This latter degree of freedom can be expressed using [10]: $\omega(\phi,t) = \sum_{n=2}^{k_{tot}} A_n(t)\cos n\phi + B_n(t)\sin n\phi$, in which the rigid body motion has been eliminated. In order to generate as simple a model as possible, only one mode shape, the $n^{th}$ one, will be considered for the ring, hence: $\omega(\phi,t) = A_n(t)\cos n\phi + B_n(t)\sin n\phi$. Moreover, the considered ring is assumed to be inextensible, thus implying that its radial displacement can be expressed from its tangential displacement by: $u_s(\phi,t) = \frac{\partial\omega(\phi,t)}{\partial\phi}$. The beam free ends are assumed to remain in steady-state contact with the inner surface of the ring, therefore, a link relationship between the pertinent degrees of freedom must be written as follows: $u_s(\phi=\phi_j,t) = -u_{t_j}(x=R_{stat},t)\cos\alpha_j + \upsilon_{f_j}(x=R_{stat},t)\sin\alpha_j$, with $\alpha_j$ being the angle between the $j^{th}$ beam and the normal to the ring's inner surface. Since an energy method has been applied to develop the entire model, the kinetic energy and potential energy are defined for both the beams and the ring. Rubbing strength is introduced by defining its work. The expressions of these energies and of this work are given in Appendix A, along with expressions for the mass matrix, stiffness matrix, circulatory matrix and gyroscopic matrix associated with this model. To better understand the phenomenon appearing within this structure however, the beams are first considered to be normal to the ring's inner surface ($\alpha_j = 0°$). In this case, the system can be separated into simpler structures. The first such structure consists of beams with just a traction/compression degree of freedom rubbing on the ring. The second consists of beams with just a flexural motion rubbing on the ring. Then, both of these degrees of freedom will be combined. In these cases and for the sake of simplicity (to handle modal mass and stiffness), the beams will be compared to radial spring-masses having two degrees of freedom (see Fig. 1b). The associated model has been developed in Appendix B. Lastly, the effect of an angle of inclination between the beams and the ring will be analysed.

## III. ROTATING RADIAL BEAMS RUBBING ON A FLEXIBLE RING

The stability of an elastic ring rubbed by one or several beams can be investigated by determining the solution $\lambda=a+ib$ to the characteristic equation $\det(\lambda^2\mathbf{M}+\lambda(\mathbf{G}+\mathbf{R})+\mathbf{K})=0$, where $\mathbf{M}$, $\mathbf{G}$, $\mathbf{R}$ and $\mathbf{K}$ are the mass matrix, gyroscopic matrix, circulatory matrix and stiffness matrix of the system, respectively. The system becomes unstable if one or more of the eigenvalue real parts $a$ are positive. Throughout this section, the beams are assumed to be radial to the ring's inner surface ($\alpha_j=0°$).

### 1. BEAMS WITH JUST A TRACTION/COMPRESSION DEGREE OF FREEDOM

In this section, the beams considered contain only a traction/compression degree of freedom. They can thus be represented by radial rotating spring-masses rubbing on the elastic ring, as plotted on Figure 2 in the particular case of just one rotating load. Due to the link relationship between the radial degrees of freedom of the model, the system has two degrees of freedom and the associated matrices can be deduced from the complete system developed in Appendix B. In the case of just one rotating load, the dynamic behaviour of the system can thus be described by the following matrix equation:

$$\begin{bmatrix} M_{stat}\left(n^2+1\right) & -\mu\left\{1+\frac{h}{2R_{stat}}\left(n^2-1\right)\right\}m_r n \\ 0 & M_{stat}\left(n^2+1\right)+m_r n^2 \end{bmatrix}\begin{Bmatrix}\ddot{A}_n\\ \ddot{B}_n\end{Bmatrix}+\begin{bmatrix} 0 & -2M_{stat}n\Omega\left(n^2+1\right) \\ 2M_{stat}n\Omega\left(n^2+1\right) & 0\end{bmatrix}\begin{Bmatrix}\dot{A}_n\\ \dot{B}_n\end{Bmatrix}$$

$$+\begin{bmatrix} K_{stat}n^2(n^2-1)^2-M_{stat}n^2\Omega^2\left(n^2+1\right) & -\mu\left\{1+\frac{h}{2R_{stat}}\left(n^2-1\right)\right\}\left(k_r-m_r\Omega^2\right)n \\ 0 & K_{stat}n^2(n^2-1)^2-M_{stat}n^2\Omega^2\left(n^2+1\right)+(k_r-m_r\Omega^2)n^2 \end{bmatrix}\begin{Bmatrix}A_n\\ B_n\end{Bmatrix}=\begin{Bmatrix}\mu\left\{1+\frac{h}{2R_{stat}}\left(n^2-1\right)\right\}\left(m_r\Omega^2R_{stat}+N_U\right)\\ -m_rR_{stat}\Omega^2 n\end{Bmatrix} \quad (1)$$

$$\text{with } M_{stat}=\rho_{stat}S_{stat}R_{stat}\pi \text{ and } K_{stat}=\frac{E_{stat}I_{stat}\pi}{R_{stat}^3}$$

It is obvious that rubbing makes the mass and stiffness matrices asymmetric, which is known to be characteristic of a potentially-unstable system. Some potential critical speeds of the system may be determined analytically using the Routh-Hurwitz criterion. The characteristic polynomial of this matrix equation actually has the following form: $P(s)=As^4+Bs^3+Cs^2+Ds+E$ with:

$$A=\left[M_{stat}\left(n^2+1\right)\right]^2+M_{stat}\left(n^2+1\right)n^2m_r$$

$$B=\left\{1+\frac{h}{2R_{stat}}\left(n^2-1\right)\right\}\mu m_r 2M_{stat}n^2\left(n^2+1\right)\Omega$$

$$C=K_{stat}n^2(n^2-1)^2\left[2M_{stat}\left(n^2+1\right)+n^2m_r\right]+M_{stat}n^2\Omega^2\left(n^2+1\right)^2\left[2M_{stat}-m_r\right]+M_{stat}\left(n^2+1\right)n^2k_r$$

(2)

$$D=\left\{1+\frac{h}{2R_{stat}}\left(n^2-1\right)\right\}\left[k_r-m_r\Omega^2\right]\mu 2M_{stat}n^2\left(n^2+1\right)\Omega$$

$$E=\left[-M_{stat}n^2\Omega^2\left(n^2+1\right)+K_{stat}n^2(n^2-1)^2\right]\left[-n^2\Omega^2\left(M_{stat}\left(n^2+1\right)+m_r\right)+K_{stat}n^2(n^2-1)^2+n^2k_r\right]$$

According to the Routh-Hurwitz criterion, the polynomial $P(s)=As^4+Bs^3+Cs^2+Ds+E$ has all its roots with real parts negative if $A$, $B$, $\frac{BC-AD}{B}$, $\frac{\left(BC-AD\right)D-B^2E}{BC-AD}$ and $E$ have the same sign. Each sign change of one of these terms implies that one of the roots of the characteristic polynomial crosses the vertical axis, making its real part positive, hence the system becomes unstable. It is obvious that $A$ and $B$ are always positive.

Regarding the term $\frac{BC-AD}{B}$:

- If $M_{stat}\left(n^2+1\right)\left(1+2n^2\right) > m_r n^4$, it is positive if:

$$\Omega^2 > \Omega_{c4}^2 = \omega_r^2 \frac{M_{stat}\left(n^2+1\right)}{\left[M_{stat}\left(n^2+1\right)\left(1+2n^2\right)-m_r n^4\right]} - \Omega_c^2 \frac{n^2\left[m_r n^2 + 2M_{stat}\left(n^2+1\right)\right]}{\left[M_{stat}\left(n^2+1\right)\left(1+2n^2\right)-m_r n^4\right]}$$

provided that $\omega_r^2 > \Omega_c^2 \dfrac{n^2\left[m_r n^2 + 2M_{stat}\left(n^2+1\right)\right]}{M_{stat}\left(n^2+1\right)}$, otherwise $\Omega_{c_4} = 0$.

- If $M_{stat}\left(n^2+1\right)\left(1+2n^2\right) < m_r n^4$, it is negative if $\Omega^2 > \Omega_{c4}^2$ provided that :

$\omega_r^2 < \Omega_c^2 \dfrac{n^2\left[m_r n^2 + 2M_{stat}\left(n^2+1\right)\right]}{M_{stat}\left(n^2+1\right)}$, otherwise $\Omega_{c_4} = 0$.

Regarding the term $\dfrac{\left(BC-AD\right)D - B^2E}{BC-AD}$:

- If $\omega_r^2 > \Omega_c^2 \dfrac{n^2\left[m_r n^2 + 2M_{stat}\left(n^2+1\right)\right]}{M_{stat}\left(n^2+1\right)}$: the numerator is positive if

$$\Omega_{c5_1}^2 = \frac{-\left[2\omega_r^2\left(n^2+1\right)+2\Omega_c^2 n^2\left(n^2-1\right)\right]+4n^2\Omega_c^2\sqrt{\left(n^2+1\right)\omega_r^2 - n^2\Omega_c^2}}{-2\left(n^2+1\right)^2} < \Omega^2 < \Omega_{c5_2}^2 = \frac{-\left[2\omega_r^2\left(n^2+1\right)+2\Omega_c^2 n^2\left(n^2-1\right)\right]-4n^2\Omega_c^2\sqrt{\left(n^2+1\right)\omega_r^2 - n^2\Omega_c^2}}{-2\left(n^2+1\right)^2}$$

provided that $-\left[2\omega_r^2\left(n^2+1\right)+2\Omega_c^2 n^2\left(n^2-1\right)\right]+4n^2\Omega_c^2\sqrt{\left(n^2+1\right)\omega_r^2 - n^2\Omega_c^2} < 0$ ; otherwise, if $0 < \Omega^2 < \Omega_{c5_2}^2 = \dfrac{-\left[2\omega_r^2\left(n^2+1\right)+2\Omega_c^2 n^2\left(n^2-1\right)\right]-4n^2\Omega_c^2\sqrt{\left(n^2+1\right)\omega_r^2 - n^2\Omega_c^2}}{-2\left(n^2+1\right)^2}$. The denominator is positive for $\Omega_c^2 > \Omega_{c4}^2$ if $M_{stat}\left(n^2+1\right)\left(1+2n^2\right) > m_r n^4$; otherwise, it is always negative.

- If $\Omega_c^2 \dfrac{n^2\left[m_r n^2 + 2M_{stat}\left(n^2+1\right)\right]}{M_{stat}\left(n^2+1\right)} > \omega_r^2 > \Omega_c^2 \dfrac{n^2}{\left(n^2+1\right)}$: the numerator is positive if

$$\Omega_{c5_1}^2 = \frac{-\left[2\omega_r^2\left(n^2+1\right)+2\Omega_c^2 n^2\left(n^2-1\right)\right]+4n^2\Omega_c^2\sqrt{\left(n^2+1\right)\omega_r^2 - n^2\Omega_c^2}}{-2\left(n^2+1\right)^2} < \Omega^2 < \Omega_{c5_2}^2 = \frac{-\left[2\omega_r^2\left(n^2+1\right)+2\Omega_c^2 n^2\left(n^2-1\right)\right]-4n^2\Omega_c^2\sqrt{\left(n^2+1\right)\omega_r^2 - n^2\Omega_c^2}}{-2\left(n^2+1\right)^2}$$

and $-\left[2\omega_r^2\left(n^2+1\right)+2\Omega_c^2 n^2\left(n^2-1\right)\right]+4n^2\Omega_c^2\sqrt{\left(n^2+1\right)\omega_r^2 - n^2\Omega_c^2} < 0$ ; otherwise, if

$$0 < \Omega^2 < \Omega_{c5_2}^2 = \frac{-\left[2\omega_r^2\left(n^2+1\right)+2\Omega_c^2 n^2\left(n^2-1\right)\right]-4n^2\Omega_c^2\sqrt{\left(n^2+1\right)\omega_r^2 - n^2\Omega_c^2}}{-2\left(n^2+1\right)^2}.$$

The denominator is negative for $\Omega_c^2 > \Omega_{c4}^2$ if $M_{stat}\left(n^2+1\right)\left(1+2n^2\right) < m_r n^4$; otherwise, it is always positive.

- If $\Omega_c^2 \dfrac{n^2}{\left(n^2+1\right)} > \omega_r^2$ the numerator is always negative. The sign of the denominator is the same as in the above case.

The last term $E$ is negative between $\Omega_{c2}$ and $\Omega_c$ with $\Omega_{c2}^2 = \dfrac{\Omega_c^2}{1+\dfrac{m}{M_{stat}\left(n^2+1\right)}} + \dfrac{\omega^2}{1+\dfrac{M_{stat}\left(n^2+1\right)}{m}}$ and $\Omega_c^2 = \dfrac{K_{stat}}{M_{stat}}\dfrac{\left(n^2-1\right)^2}{n^2+1}$, corresponding to the ring's first critical speed. These last two rotational speeds determine the rotational speed range over which the system is unstable even without rubbing. It will be shown below that this instability consists of a divergence in the ring's forward mode shape. This phenomenon is close to that shown by Canchi and Parker [8] or by Iwan and Stahl [3], and Iwan and Moeller [4] in the case of a disk instead of a ring, with the influences of the load parameters also being quite similar.

In all these expressions, $\omega_{r_j}^2 = \frac{k_{r_j}}{m_{r_j}}$ is the squared angular frequency of the radial spring-mass.

It can thus be seen that this kind of system with rubbing is almost always unstable. As a matter of fact, it only lies within specific rotational speed ranges, i.e. only between $\Omega_{c_{5_1}}$ and $\Omega_{c_{5_2}}$ can the above coefficients all be positive in the case of a lightweight system in comparison with the ring's (i.e. $m_r n^4 < M_{stat}\left(n^2+1\right)\left(1+2n^2\right)$) and a stiffness, such that: $\omega_r^2 > \Omega_c^2 \frac{n^2}{\left(n^2+1\right)}$.

The effects of both a mass rubbing on the ring and of the stiffness may be separated. Figures 3a and 3c display the stability analysis of a radial stiffness (without mass) rubbing against the ring with $\mu = 0.01$ and $\mu = 0.1$, respectively. Figures 3b and 3d show associated zooms of Figs. 3a and 3c, respectively. As explained previously, a divergence instability in the forward mode shape of the ring between $\Omega_c$ and $\Omega_{c2}$ can be observed. Moreover, as expected, once the rotational speed is greater than 0 RPM, the system, and especially the backward mode shape of the ring, is unstable because of rubbing. This rubbing effect is well-known and has been highlighted, for instance, in the case of a modal representation of a turbine engine excited by rubbing forces [11]. It thus appears that as the rubbing coefficient increases, instability rises even faster. The case of just one mass rubbing on a ring will now be considered; Figure 4 presents the associated stability analysis. Here again, the system is unstable as soon as the rotational speed differs from zero, with the unstable mode now however being the ring's forward mode shape. A divergence instability in the forward mode shape nonetheless remains after reaching the critical system speed, between $\Omega_{c_2}$ and $\Omega_c$, and mode coupling between the forward and backward mode shapes of the ring. It should be noted that this mode-coupling appears even without rubbing and is due to load displacement on the elastic ring. This phenomenon has been reported in the case of a disk instead of a ring (see [3,4]). Figures 4c and 4d show the stability analysis for the same system as in Figures 4a and 4b, but with a higher rubbing coefficient. The effect of the rubbing coefficient is the same as before. Both a stiffness and mass will now be considered. Figures 5a and 5c (with the associated zooms in Figs. 5b and 5d, respectively) display stability analysis for a radial spring-mass rubbing on the ring's two-node diameter mode shape. In both cases, the mass is $m_r = 100kg$, yet Figures 5a and 5b include $k_r = 1.10^6 N.m^{-1}$, whereas $k_r = 1.10^5 N.m^{-1}$ in Figures 5c and 5d. A cross between the real part curves of the ring's forward and backward mode shapes can be observed, which seems to be correct in comparison with the last results taken separately. At low rotational speeds, the stiffness actually destabilises the backward mode shape of the ring, but at higher rotational speeds the mass, with a negative stiffening effect proportional to rotational speed ($-m_r\Omega^2$), destabilises the ring's forward mode shape. This cross only occurs if $\omega_r > \omega_n$, where $\omega_r$ is the angular frequency of the spring-mass and $\omega_n$ the angular frequency of the ring's $n^{th}$ nodal diameter mode shape. Moreover, as indicated on these last figures, the cross occurs at a rotational speed between $\Omega_{c_{5_1}}$ and $\Omega_{c_{5_2}}$. As earlier discussed, the system may be stable between $\Omega_{c_{5_1}}$ and $\Omega_{c_{5_2}}$, as shown in Figures 5b and 5d. Figure 6 presents a stability analysis for the same system as in Figure 5a, but with a higher rubbing coefficient, once again emphasising its effect. It can be pointed out that the rubbing coefficient exerts no effect on the remarkable critical rotational speeds $\Omega_c$, $\Omega_{c_2}$, $\Omega_{c_{5_1}}$ and $\Omega_{c_{5_2}}$. As the number of nodal diameters of the ring's mode shape increases to infinity, the speed range $\left[\Omega_{c_{5_1}}, \Omega_{c_{5_2}}\right]$ collapses to $\Omega_c$, which itself tends to infinity.

The case of several radial rotating spring-masses rubbing on the ring will now be investigated. Certain configurations appear to avoid the divergence instability of the forward mode shape between $\Omega_c$ and $\Omega_{c_2}$, as shown in Figure 7 in the case of spring-masses with $\omega_r = 100\, rad/s$ and $\mu = 0.01$.

Since this divergence instability is present without rubbing, the Routh-Hurwitz criterion applied to the characteristic polynomial of the system with $\mu = 0$ can yield a sufficient condition for the disappearance of divergence. This condition may be written as:

$$\begin{cases} \sum_j k_{r_j} \sin^2(n\phi_j) = \sum_j k_{r_j} \cos^2(n\phi_j) \\ \sum_j k_{r_j} \sin(n\phi_j)\cos(n\phi_j) = 0 \\ \sum_j m_{r_j} \sin^2(n\phi_j) = \sum_j m_{r_j} \cos^2(n\phi_j) \\ \sum_j m_{r_j} \sin(n\phi_j)\cos(n\phi_j) = 0 \end{cases} \quad (3)$$

These conditions are obviously satisfied in the case shown in Figure 7c since all the spring-masses have the same parameter values and are located at $\phi_1 = 60°$, $\phi_2 = 120°$ and $\phi_3 = 180°$, which is not true in the cases shown on the other figures. It can be noted that even in the case with no divergence of the ring's forward mode shape, the system is still unstable once the rotational speed differs from 0 RPM.

## 2. BEAMS WITH JUST ONE FLEXURAL DEGREE OF FREEDOM

Here again, this system is quite similar to a rubbing rotating spring-mass tangent to the ring, as depicted in Figure 8, in the case of one spring-mass. The matrix equation for the dynamic behaviour of such systems is now: $(2 + \textit{number of loads}) \times (2 + \textit{number of loads})$. From a stability analysis point of view, the differences between the beam model with just one flexural degree of freedom and the tangent spring-mass model stem from the spin-softening terms. Those associated with the beam model do not take into account the entire flexural modal mass, but rather $\left( m_{t_j} - \rho_{b_j} I_{b_j} \frac{\pi^2}{8R_{stat}} \right)$. Another difference also concerns matrix $\mathbf{R}$, which is neither symmetric nor skew-symmetric (see Appendices A and B). The phenomenon occurring for the tangent spring-masses rubbing on the ring should however be the same as for beams with just a flexural degree of freedom rubbing on a ring. The stability analysis of such systems can thus be performed using the tangent spring-mass model, which allows considering load modal parameters. In the case of only one tangent spring-mass rubbing on the casing, the characteristic polynomial of its matrix equation is:

$$P(s) = \left[ s^2 m_t + s\left(2\mu m_t \Omega\right) + k_t - m_t \Omega^2 \right] \left[ \left( s^2 \left( \rho M_{stat} \left(n^2+1\right)\right) + K_{stat} n^2 (n^2-1)^2 - M_{stat} n^2 \Omega^2 \left(n^2+1\right) \right)^2 + s^2 \left( 2 M_{stat} n \Omega \left(n^2+1\right)\right)^2 \right]$$

By calculating just the roots of this polynomial, which correspond with the roots of its first member $s^2 m_t + s\left(2\mu m_t \Omega\right) + k_t - m_t \Omega^2$, spring-mass stability can be studied. The discriminant of this first member is: $\Delta = \Omega^2 \left(2m_t\right)^2 \left(\mu^2+1\right) - 4 m_t k_t$. If $\Omega^2 < \frac{k_t}{m_t\left(\mu^2+1\right)}$, then $\Delta < 0$ and the roots of this polynomial are:

$$s_1 = \frac{-2\mu m_t \Omega + i\sqrt{\Delta}}{2m_t} \quad and \quad s_2 = \frac{-2\mu m_t \Omega - i\sqrt{\Delta}}{2m_t},$$

thus $\mathrm{Re}(s_1) < 0$ *and* $\mathrm{Re}(s_2) < 0$ and the spring-mass is stable. Now, if $\Omega^2 > \frac{k_t}{m_t\left(\mu^2+1\right)}$, then $\Delta > 0$ and $s_1 = \frac{-2\mu m_t \Omega + \sqrt{\Delta}}{2m_t}$ *and* $s_2 = \frac{-2\mu m_t \Omega - \sqrt{\Delta}}{2m_t}$, thus $\mathrm{Re}(s_2) < 0$. Concerning the real part of $s_1$, it is negative if $\Omega^2 < \frac{k_t}{m_t} = \omega_t^2$ and positive if $\Omega^2 > \omega_t^2$, corresponding to a system divergence. All these results are valid for the beam, but the remarkable rotational speeds are $\frac{4m_t k_t}{\left(2\mu\rho_b S_b \frac{R_{stat}}{\pi}\right)^2 + 4m_t\left(m_t - \rho_b I_b \frac{\pi^2}{8R_{stat}}\right)}$ instead of $\frac{k_t}{m_t\left(\mu^2+1\right)}$ and $\frac{k_t}{\left(m_t - \rho_b I_b \frac{\pi^2}{8R_{stat}}\right)}$ instead of $\frac{k_t}{m_t}$. Figure 9 presents the stability analysis of one tangent spring-mass rubbing on the ring's two-node diameter mode shape, with $k_t = 1.10^6\, N.m^{-1}$ and $m_t = 100kg$, and with: a) $\mu = 0.01$ and b) $\mu = 0.1$. As expected, the spring-mass is stable until $\Omega^2 > \omega_t^2$, at which point it experiences divergence instability. The effect of the rubbing coefficient is the same as before. In the case of several tangent spring-masses rubbing against the ring, no additional phenomenon occurs. It can also be observed that both of the ring's mode shapes appear to be perfectly stable.

Since the effects of each degree of freedom for a beam rubbing on an elastic ring have been studied separately, the beams can now be considered to possess both degrees of freedom.

## 3. BEAMS WITH BOTH A TRACTION/COMPRESSION DEGREE OF FREEDOM AND A FLEXURAL DEGREE OF FREEDOM

The beam's two degrees of freedom will now be considered. Once again, this system, as detailed in Appendix A, is similar to a spring-mass with two degrees of freedom (see Appendix B), as displayed in Figure 1. The differences between these two models, in addition to all those described above, stem from the gyroscopic terms present since the spring-masses have two degrees of freedom that are not expressed in the same manner. These gyroscopic terms are likely to create new mode couplings in the system. Nevertheless, as seen in the latter case, the phenomenon appearing for these two systems should be the same. The spring-mass system will be studied in order to easily handle modal parameters and afterwards will be compared with the beam model.

Figure 10 shows the stability analysis of the two-node diameter mode shape of the ring rubbed by a spring-mass with $m = 100kg$, $k_r = k_t = 1.10^6\, N.m^{-1}$ and $\mu = 0.1$. All phenomena studied earlier resulting from a tangent spring-mass or a radial spring-mass rubbing on the ring are once again present. The effect of the rubbing coefficient (not represented here) is still the same: an increase in the slope of the curves' real part. A locus veering phenomenon is also in effect between the ring's backward mode shape and the spring-mass, followed by mode coupling between the ring's forward mode shape and the spring-mass. This mode coupling may result from the gyroscopic terms. Moreover, the speed range concerned by this mode coupling is very sensitive to the tangential stiffness $k_t$ of the spring-mass, as shown in Figure 11. The greater the tangential stiffness, the wider the range in mode coupling speed.

Lastly, the effects of several rotating spring-masses can also be studied. Figure 12 shows the stability analysis for the two-node diameter mode shape of the ring rubbed by two identical spring-masses with $m = 100kg$, $k_r = k_t = 1.10^6\, N.m^{-1}$ and $\mu = 0.1$; Figure 12a corresponds to two loads separated by 60° from each other, whereas Figure 12b corresponds to two loads separated by 180°. In both cases, it appears that only one spring-mass exchanges its mode shape with the backward mode shape of the ring (locus veering) and then experiences mode coupling with this ring's forward mode shape. Moreover, in the first case (i.e. spring-masses separated by 60°), the eigenfrequencies of both spring-masses slightly increase after their theoretical divergence, as shown in Figure 12a; this does not occur when the two spring-masses are diametrically opposed (see Fig. 12b). This phenomenon will be analysed further below. Figure 13 exhibits the stability analysis for the two-node diameter mode shape of the ring rubbed by three identical rotating spring-masses with $m = 100kg$, $k_r = k_t = 1.10^6\, N.m^{-1}$ and $\mu = 0.1$, but either separated by 60° from each other (Fig. 13a), or two separated by 60° and the third at 180° from one of the other two (Fig. 13b). In the first case, the sufficient condition for eliminating the divergence instability between $\Omega_c$ and $\Omega_{c_2}$ is satisfied, hence Figure 13a shows no divergence between these rotational speeds, whereas this divergence is still present in Figure 13b. Once again, in both cases, only one spring-mass exchanges its mode shape with the ring's backward mode shape and then experiences mode coupling with this ring's forward mode shape. Moreover, after the theoretical divergence in rotational speed for the three spring-masses, two of them also seem to have slightly increased in eigenfrequency. The spring-mass eigenfrequencies can be adjusted through their masses and stiffness. A stability analysis for the two-node diameter mode shape of the ring rubbed by two spring-masses with two different eigenfrequencies separated by 60° from each other has been plotted on Figure 14. In this case, just one spring-mass exchanges its mode shape with the ring's backward shape, yet both spring-masses experience mode coupling with the ring's forward mode shape. The system, like in all other examples, is unstable once the rotational speed differs from zero.

Concerning the increase in eigenfrequencies of the divergent beams (see Figs. 12a and 13), there is actually a transition between the divergence and flutter of two, and only two, beams becoming coupled through the ring. The mode shape being considered for the ring is indeed very important for this coupling between two beams and the ring that provides steady-state contact. Although all simulations presented in this paper pertain to the ring's two-node diameter mode shape, Figure 15 shows the angular regions where a beam located at 60° in the rotating frame can couple with another beam for: a) the ring's two-node diameter mode shape, and b) its three-node diameter mode shape. It thus appears that four regions exist for the ring's two-node diameter mode shape, whereas six exist with the ring's three-node diameter mode shape. It can moreover be seen that as ring deformation increases, the coupling regions become narrower. For the two-node diameter mode shape for example, each coupling region is 36° wide, while for the three-node diameter mode shape, each one is 16° wide. It can be noted that the whole angular position range, over which two blades can couple, is greater in the case of the ring's two-node diameter mode shape (144°) than in the case of its three-node diameter mode shape (96°). Figure 16 shows a zoom of Figure 12a near the coupling region. It can clearly be seen on this image that after beam divergence, the associated real parts couple with one another, thereby leading to an unstable dynamic configuration. A case of changing instability has been uncovered by Gaul and Wagner [12], whereby a rotating system experienced instability divergent from mode-coupling. Moreover, the rotational speed at which coupling appears varies over the coupling angular region, as indicated in Table 1 for the case of the two-node diameter mode shape of the ring rubbed ($\mu = 0.1$) by two spring-masses with $m = 100kg$ and $k_r = k_t = 1.10^6 N.m^{-1}$, the first one being at 60° in the rotating frame and the other between 87° and 123°. It must be pointed out that this phenomenon also occurs with beams featuring different modal parameters.

It has been said prior that phenomena occurring in the case of a spring-mass with two degrees of freedom rubbing on the flexible ring are the same as those occurring in the case of a beam also with two degrees of freedom rubbing on this ring. This can be confirmed for one rotating load rubbing on the ring's two-node diameter mode shape. Figure 17 displays a stability analysis for this ring's mode shape rubbed either by a spring-mass (Fig. 17a) or by a beam (Fig. 17b), both having the same modal parameters: $\omega_r = 251\, rad/s$ and $\omega_t = 100\, rad/s$ (for the spring-mass: $m = 142.8kg$, $k_r = 9.036.10^6 N.m^{-1}$ and $k_t = 1.428.10^6 N.m^{-1}$ and for the beam: $m_r = 185.66kg$, $m_t = 100kg$, $k_r = 1.175.10^7 N.m^{-1}$ and $k_t = 1.10^6 N.m^{-1}$). As expected, the same phenomena occur in both cases: locus veering is in effect between the spring-mass and the ring's backward mode shape, along with a divergence in its forward mode shape between $\Omega_c$ and $\Omega_{c_2}$, mode coupling between this forward mode shape and the spring-mass, a divergence in this spring-mass and mode coupling between the ring's forward and backward mode shapes. The only difference between these two systems is the offset of these phenomena due to modal parameter differences. The system naturally becomes unstable once the rotational speed differs from 0 RPM.

# IV. ROTATING BEAMS RUBBING ON A FLEXIBLE RING WITH AN ANGLE OF INCLINATION

The effect of an angle of inclination of a beam rubbing on a rotating disk has been studied by, among others, Chambrette and Jezequel [7], yet no studies have been found in the literature on the influence of such an angle between beams rotating with rubbing on the inner surface of an elastic ring. The previous study by Chambrette and Jezequel [7] demonstrated that this kind of angle can modify the parametric domains where the system is unstable: the investigated system was a rotating disk excited by a beam with both traction/compression motion and flexural motion. It has been shown that the same kind of instabilities as those included in the present study, i.e. divergence after the critical speed and mode coupling, could be obtained and modified by the angle between the beam and the disk. The main results from this study were in fact that as the beam became more heavily inclined, the ring's divergence speed range narrowed and the ring's mode coupling was more heavily delayed. It has also been reported that mode couplings could arise even before the critical speed if the frequency of the beam's flexural degree of freedom was below the disk's frequency or close to it. The influence of the angle of inclination on the present system, in the case of just one beam rubbing on the ring, could thus be studied when the frequency of the beam's flexural motion lies below or above or close to that of the ring. It appears however that the mechanisms occurring due to this angle of inclination are the same in all three of these cases; therefore, only the case where the beam's flexural motion frequencies lie below ring frequencies will be detailed herein.

The ring's two-node diameter mode shape has been set at 30 Hz and the flexural motion of the beam at 20 Hz. Figure 18 presents the stability analysis for a beam rubbing on the ring's two-node diameter mode shape, with: a) $\alpha = 0°$, b) $\alpha = 5°$, c) $\alpha = 10°$ and d) $\alpha = 89°$. These values of $\alpha$ have been chosen because of the high evolution in system frequencies for low values of $\alpha$. Figure 19 exhibits the associated appropriate zooms of Figure 18. First of all, it may be observed on Figure 18 that as $\alpha$ increases, the rotational speed at which the system experiences mode coupling between the ring's forward and backward mode shapes rises, as does the rotational speed at which the beam's flexural degree of freedom diverges. Moreover, mode coupling between the ring's forward mode shape and this flexural degree of freedom of the beam appears before divergent instability in the latter, when $\alpha$ increases, as shown on Figure 19. On this same figure, it appears that the rotational speed range over which the ring's forward mode shape diverges decreases as $\alpha$ increases and then finally disappears. For low values of $\alpha$, a locus veering phenomenon between the ring's forward mode shape and the beam's flexural motion may be observed, especially on Figure 19a. During this phenomenon, the system is bound to be unstable; this instability (mode coupling type) can thus take effect before the ring's critical speed. As $\alpha$ increases, the flexural degree of freedom frequency increases and hence the locus veering phenomenon with the forward mode shape disappears. As for mode coupling between the ring's forward mode shape and the beam's flexural motion mentioned above, Figure 19 shows that as $\alpha$ increases, the associated rotational speed range begins later and has greater values. This instability occurs after the ring's critical speed. All these phenomena may be seen continuously as a function of $\alpha$, as indicated on Figures 20 and 21. The evolution in the rotational speed at which both mode shapes of the ring couple can be monitored on Figure 20. Figure 21 shows the evolution of the post-critical mode couplings between the ring's forward mode shape and the beam's flexural motion, as well as the divergence of this latter degree of freedom. The rotational speed range associated with this mode coupling may be increased about 500 RPM, from a configuration at $\alpha = 0°$ to one at $\alpha = 89°$.

The mechanism involved in the inclination of a beam rubbing on an elastic ring thus primarily consists of an increase in the beam's flexural frequency. If, when the beam is radial to the ring, its flexural frequency is below that of the ring because of the evolution with rotational speed, either locus veering or mode coupling can occur between the ring's forward mode shape and the beam's flexural motion even before the ring's critical speed. In this case, both eigenfrequencies (of the beam and the ring) are very close to each other (see Fig. 19a), and the system is bound to be unstable. As $\alpha$ increases, beam frequencies increase until reaching the frequency of its traction/compression degree of freedom. Once the beam frequency has risen above ring frequencies (from $\alpha > 15°$), locus veering concerns its backward mode shape and, as seen on Figure 19d, the specific eigenfrequencies are no longer close to one another. This veering does not cause system instability.

This mechanism is the same as in the case where the beam's flexural frequency is higher than that of the ring. As $\alpha$ increases, the beam's flexural frequency increases; however, since it always remains above the ring's, locus veering may occur even before the ring's critical speed yet can only concern its backward mode shape. In this case, both frequencies are not very close to one another and this veering does not make the system unstable. While the beam's flexural frequency decreases with an increase in rotational speed, mode coupling with the ring's forward mode shape then occurs.

The influence of the angle of inclination of the beam rubbing on an elastic ring is therefore close to that of a beam rubbing on a disk (see [7]). This angle acts upon the same critical phenomena. As the inclination angle increases, the rotational speed range over which the ring's forward mode shape diverges can in fact be modified (reduced), as can the rotational speed for both mode shapes of the ring couple (put away). The system can also be made unstable before the ring's critical speed by means of mode coupling between the beam's flexural motion and the ring's forward mode shape provided the flexural frequency lies below the ring's frequency. The beam's angle of inclination actually modifies the values of the normal and tangential strength between both structures in contact, thereby modifying phenomena like divergence or the ring's mode coupling. This angle also modifies the flexural frequency of the beam in contact with the ring, making mode couplings possible or not provided items have frequencies close to each other.

All simulations have been conducted for a ring's two-node diameter mode shape, yet the same phenomena are present for other mode shapes as well. Moreover, only one mode shape for the ring and beams has been considered herein; the phenomena targeted in this study however are quite similar to those that may arise when considering several mode shapes for each item of the model, as illustrated by Iwan and Stahl [3] and Iwan and Moeller [4].

## V. CONCLUSION

The stability of rotating beams rubbing on an elastic ring has been studied in this article. An energy model of flexible beams possessing two degrees of freedom in steady-state contact with an elastic ring possessing just one in-plane mode shape has been developed within the rotating frame. This model, devoid of time-dependent terms, has been studied from a stability point of view. It appears that rubbing always makes the system unstable once the beam's rotational speed is nonzero. It has also been shown that a radial stiffness rubbing on the ring tends to make its backward mode shape unstable, whereas a concentrated mass rubbing on a ring makes the forward mode shape unstable. The traction/compression degree of freedom of a beam rubbing on a ring, in addition to the unstable phenomena occurring even without rubbing (divergence of the ring's forward mode shape near its critical speed and post-critical mode coupling between forward and backward mode shapes), thus starts by making its backward mode shape unstable and then its forward mode shape. The remarkable rotational speeds of these phenomena have been determined analytically. As the rubbing coefficient rises, the gradient of the eigenvalue real parts also rises. The beams' flexural degree of freedom yields mode couplings and locus veering with the ring. The influence of several beams rubbing on a ring has been examined and some cases of coupling between beams highlighted. Lastly, an angle of inclination between the beams and the ring has been considered. It has also been demonstrated that the main result of this parameter was the increase in the beam's flexural frequency with inclination, thus leading to veering and mode couplings.

APPENDIX A

Expressions of the kinetic energy and potential energy, as well as the work of rubbing strength associated with the model of $j$ rotating beams located at $\phi_j$ within the rotating frame rubbing on the flexible inextensible ring with inclination angle $\alpha_j$.

The expression of the kinetic energy of the system is given by:

$$T=\frac{1}{2}\int_{-\Omega t}^{2\pi-\Omega t}\rho_{stat}S_{stat}\left\{\left[\dot{u}_s(\phi,t)-\Omega\frac{\partial u_s}{\partial\phi}(\phi,t)\right]^2+\left[\dot{w}(\phi,t)-\Omega\frac{\partial w}{\partial\phi}(\phi,t)\right]^2\right\}R_{stat}d\phi$$

$$+\sum_j\frac{1}{2}\int_0^{R_{stat}}\rho_{b_j}S_{b_j}\left\{\dot{u}_{t_j}^2(x,t)+\dot{\upsilon}_{f_j}^2(x,t)+\Omega^2\left(\left(x+u_{t_j}(x,t)\right)^2+\upsilon_{f_j}^2(x,t)\right)+2\Omega\left(\dot{\upsilon}_{f_j}(x,t)\left(x+u_{t_j}(x,t)\right)-\dot{u}_{t_j}(x,t)\upsilon_{f_j}(x,t)\right)\right\}dx$$

$$+\sum_j\frac{1}{2}\int_0^{R_{stat}}\rho_{b_j}I_{b_j}\left(\Omega+\frac{\partial\dot{\upsilon}_{f_j}(x,t)}{\partial x}\right)^2dx$$

The expression of the potential energy of the system is given by :

$$\gamma=\frac{1}{2}\int_{-\Omega t}^{2\pi-\Omega t}\frac{E_{stat}I_{stat}}{R_{stat}^3}\left\{\frac{\partial^2u_s}{\partial\phi^2}(\phi,t)+u_s(\phi,t)\right\}^2d\phi+\sum_j\frac{1}{2}\int_0^{R_{stat}}E_{b_j}S_{b_j}\left\{\frac{\partial u_{t_j}(x,t)}{\partial x}\right\}^2dx+\sum_j\frac{1}{2}\int_0^{R_{stat}}E_{b_j}I_{b_j}\left\{\frac{\partial^2\upsilon_{f_j}(x,t)}{\partial x^2}\right\}^2dx$$

When including Ritz functions for the degrees of freedom in the above expressions and in considering the relationship between the ring's radial degree of freedom and both the beam's degrees of freedom: $u_s(\phi=\phi_j,t)=-u_{t_j}(x=R_{stat},t)\cos\alpha_j+\upsilon_{f_j}(x=R_{stat},t)\sin\alpha_j$, these energies and potentials can be written by:

$$T=\frac{1}{2}\int_{-\Omega t}^{2\pi-\Omega t}\rho_{stat}S_{stat}\left\{\left[\dot{u}_s(\phi,t)-\Omega\frac{\partial u_s}{\partial\phi}(\phi,t)\right]^2+\left[\dot{w}(\phi,t)-\Omega\frac{\partial w}{\partial\phi}(\phi,t)\right]^2\right\}R_{stat}d\phi$$

$$+\sum_j\frac{1}{2}\rho_{b_j}S_{b_j}R_{stat}\left\{\frac{1}{2}\frac{\dot{u}_s^2(\phi_j,t)}{\cos^2\alpha_j}-\dot{u}_s(\phi_j,t)\dot{\upsilon}_{f_j}\frac{\tan\alpha_j}{\cos\alpha_j}+\Omega\frac{2\dot{u}_s(\phi_j,t)\upsilon_{f_j}}{\pi\cos\alpha_j}\right\}$$

$$+\sum_j\frac{1}{2}\rho_{b_j}S_{b_j}R_{stat}\left\{\dot{\upsilon}_{f_j}^2\left(\frac{1}{2}\tan^2\alpha_j+\left(\frac{3}{2}-\frac{4}{\pi}\right)\right)+\Omega R_{stat}\frac{\left(\pi^2-4\pi+8\right)}{\pi^2}\dot{\upsilon}_{f_j}-\Omega\frac{2u_s(\phi_j,t)\dot{\upsilon}_{f_j}}{\pi\cos\alpha_j}\right\}$$

$$+\sum_j\frac{1}{2}\rho_{b_j}S_{b_j}R_{stat}\left\{\frac{1}{2}\Omega^2\frac{u_s^2(\phi_j,t)}{\cos^2\alpha_j}-\frac{8}{\pi^2}\Omega^2R_{stat}\frac{u_s(\phi_j,t)}{\cos\alpha_j}-\Omega^2u_s(\phi_j,t)\upsilon_{f_j}\frac{\tan\alpha_j}{\cos\alpha_j}\right\}$$

$$+\sum_j\frac{1}{2}\rho_{b_j}S_{b_j}R_{stat}\left\{\Omega^2\left(\frac{1}{2}\tan^2\alpha_j+\left(\frac{3}{2}-\frac{4}{\pi}\right)\right)\upsilon_{f_j}^2+\frac{8}{\pi^2}\Omega^2R_{stat}\tan\alpha_j\upsilon_{f_j}+\frac{1}{3}\Omega^2R_{stat}^2\right\}$$

$$+\sum_j\frac{1}{2}\rho_{b_j}I_{b_j}\left(\frac{\pi^2}{8R_{stat}}\dot{\upsilon}_{f_j}^2+2\Omega\dot{\upsilon}_{f_j}+\Omega^2R_{stat}\right)$$

The expression of the potential energy of the system is now given by:

$$\gamma=\frac{1}{2}\int_{-\Omega t}^{2\pi-\Omega t}\frac{E_{stat}I_{stat}}{R_{stat}^3}\left\{\frac{\partial^2u_s}{\partial\phi^2}(\phi,t)+u_s(\phi,t)\right\}^2d\phi+\sum_j\frac{1}{2}E_{b_j}S_{b_j}\frac{\pi^2}{8R_{stat}}\frac{u_s^2(\phi_j,t)}{\cos^2\alpha_j}$$

$$+\sum_j \left\{ \frac{1}{2} E_{b_j} S_{b_j} \frac{\pi^2}{8R_{stat}} \tan^2 \alpha_j + \frac{1}{2} E_{b_j} I_{b_j} \frac{\pi^4}{32R_{stat}^3} \right\} \upsilon_{f_j}^2 - \sum_j E_{b_j} S_{b_j} \frac{\pi^2}{8R_{stat}} \frac{\tan \alpha_j}{\cos \alpha_j} u_s(\phi_j, t) \upsilon_{f_j}$$

The expression for rubbing work is given by:

$$W_{ext} = \sum_j T_{b_j \to stat} \left[ \omega(\phi,t) \left\{ 1 - \frac{h}{2R_{stat}} \right\} - \frac{h}{2R_{stat}} \frac{\partial u_s(\phi,t)}{\partial \phi} - \left[ \upsilon_{f_j}(x = R_{stat}, t) \left( \cos \alpha_j + \sin \alpha_j \tan \alpha_j \right) - u_s(\phi,t) \tan \alpha_j \right] \right] \delta(\phi - \phi_j)$$

with, in the direct centripetal frame, $T_{b_j \to stat} = -\mu N_{b_j \to stat} sign(V_{slip})$ being the rubbing strength of the $j^{th}$ beam on the ring. In this expression, $V_{slip}$ is the slip speed of the beam on the ring and $N_{b_j \to stat}$ the radial load of the $j^{th}$ beam on the stator. For instance, in the case of contacts between blades of a rotating machine and the casing, it can be expressed by the radial load due to the unbalanced mass $-N_U$ plus a dynamic load due to the dynamics of the $j^{th}$ beam. In order to include this rubbing strength into the matrix equation of system dynamic behaviour and perform a stability analysis, this rubbing strength can be expressed by:

$$T_{b_j \to stat} = \mu \left[ N_U + \frac{\rho_{b_j} S_{b_j} R_{stat}}{2} \left( \ddot{u}_s(\phi,t) - \Omega^2 u_s(\phi,t) \right) + E_{b_j} S_{b_j} \frac{\pi^2}{8R_{stat}} u_s(\phi,t) - 2\rho_{b_j} S_{b_j} \frac{R_{stat}}{\pi} \Omega \dot{u}_s(\phi,t) \tan \alpha \right] \delta(\phi - \phi_j)$$

$$+\mu \left[ 2\rho_{b_j} S_{b_j} \frac{R_{stat}}{\pi} \Omega \left( \cos \alpha_j + \sin \alpha_j \tan \alpha_j \right) \dot{\upsilon}_{f_j}(x = R_{stat}, t) - \left( \rho_{b_j} S_{b_j} \frac{R_{stat}}{2} - \rho_{b_j} S_{b_j} R_{stat} \left( \frac{3}{2} - \frac{4}{\pi} \right) - \rho_{b_j} I_{b_j} \frac{\pi^2}{8R_{stat}} \right) \ddot{\upsilon}_{f_j}(x = R_{stat}, t) \sin \alpha_j \right] \delta(\phi - \phi_j)$$

$$+\mu \left[ 4\rho_{b_j} S_{b_j} \frac{R_{stat}^2}{\pi^2} \Omega^2 - \left( E_{b_j} S_{b_j} \frac{\pi^2}{8R_{stat}} - \rho_{b_j} S_{b_j} \frac{R_{stat}}{2} \Omega^2 - E_{b_j} I_{b_j} \frac{\pi^4}{32R_{stat}^3} + \rho_{b_j} S_{b_j} R_{stat} \Omega^2 \left( \frac{3}{2} - \frac{4}{\pi} \right) \right) \upsilon_{f_j}(x = R_{stat}, t) \sin \alpha_j \right] \delta(\phi - \phi_j)$$

. This implies that rubbing strength always follows the same direction, making this model valid if the radial load due to unbalanced mass is far greater than the dynamic load due to dynamics of the $j^{th}$ beam, which is acceptable, and if $V_{slip}$ always retains the same sign. This latter condition is true for a sufficient rotational speed. In all cases, the main purpose of this model and of this study is to detect the appearance of instabilities and not to calculate potential limit cycles this far into the study. The expression of $T_{b_j \to stat}$ can be obtained by the Hamilton principle using Lagrangian multipliers. It should be noted that the Ritz functions for the beams' equal unity at their end rubbing against the ring; beam parameters appearing in this rubbing strength are hence actually modal parameters of the beams at their end rubbing against the ring. Only one mode shape of the stator has been considered at this time, i.e.: $\omega(\phi,t) = A_n(t) \cos n\phi + B_n(t) \sin n\phi$ and $u_s(\phi,t) = -nA_n(t) \sin n\phi + nB_n(t) \cos n\phi$.

The matrix equation of system dynamic behaviour is:

$$\mathbf{M\ddot{X}} + \left( \mathbf{G} + \mathbf{R} \right) \mathbf{\dot{X}} + \mathbf{KX} = \mathbf{F}$$

with: $\mathbf{X}^T = \left\{ A_n \quad B_n \quad \upsilon_{f_1} \quad \cdots \quad \cdots \quad \cdots \quad \upsilon_{f_j} \right\}$

$$\mathrm{M} = \begin{bmatrix} M_{11} & M_{12} & 0 & \cdots & \cdots & \cdots & 0 \\ M_{21} & M_{22} & 0 & \cdots & \cdots & \cdots & 0 \\ m_{r_1} \left( \frac{\tan \alpha_1}{\cos \alpha_1} - \mu \left( \cos \alpha_1 + \sin \alpha_1 \tan \alpha_1 \right) \right) n \sin(n\phi_1) & -m_{r_1} \left( \frac{\tan \alpha_1}{\cos \alpha_1} - \mu \left( \cos \alpha_1 + \sin \alpha_1 \tan \alpha_1 \right) \right) n \cos(n\phi_1) & M_{33} & 0 & \cdots & \cdots & 0 \\ \vdots & \vdots & 0 & \ddots & 0 & \cdots & 0 \\ \vdots & \vdots & \vdots & 0 & \ddots & \ddots & \vdots \\ \vdots & \vdots & \vdots & \vdots & \ddots & \ddots & 0 \\ m_{r_j} \left( \frac{\tan \alpha_j}{\cos \alpha_j} - \mu \left( \cos \alpha_j + \sin \alpha_j \tan \alpha_j \right) \right) n \sin(n\phi_j) & -m_{r_j} \left( \frac{\tan \alpha_j}{\cos \alpha_j} - \mu \left( \cos \alpha_j + \sin \alpha_j \tan \alpha_j \right) \right) n \cos(n\phi_j) & 0 & 0 & \cdots & 0 & M_{(j+2)(j+2)} \end{bmatrix}$$

$$M_{11} = M_{stat} \left( n^2 + 1 \right) + \sum_j m_{r_j} n^2 \frac{\sin^2(n\phi_j)}{\cos^2 \alpha_j} + \mu \sum_j m_{r_j} \left( \left\{ 1 + \frac{h}{2R_{stat}} \left( n^2 - 1 \right) \right\} \cos(n\phi_j) - n \sin(n\phi_j) \tan \alpha_j \right) n \sin(n\phi_j)$$

$$M_{12}=-\sum_j m_{r_j} n^2 \frac{\sin(n\phi_j)\cos(n\phi_j)}{\cos^2\alpha_j}-\mu\sum_j m_{r_j}\left(\left\{1+\frac{h}{2R_{stat}}\left(n^2-1\right)\right\}\cos(n\phi_j)-n\sin(n\phi_j)\tan\alpha_j\right)n\cos(n\phi_j)$$

$$M_{21}=-\sum_j m_{r_j} n^2 \frac{\sin(n\phi_j)\cos(n\phi_j)}{\cos^2\alpha_j}+\mu\sum_j m_{r_j}\left(\left\{1+\frac{h}{2R_{stat}}\left(n^2-1\right)\right\}\sin(n\phi_j)+n\cos(n\phi_j)\tan\alpha_j\right)n\sin(n\phi_j)$$

$$M_{22}=M_{stat}\left(n^2+1\right)+\sum_j m_{r_j} n^2 \frac{\cos^2(n\phi_j)}{\cos^2\alpha_j}-\mu\sum_j m_{r_j}\left(\left\{1+\frac{h}{2R_{stat}}\left(n^2-1\right)\right\}\sin(n\phi_j)+n\cos(n\phi_j)\tan\alpha_j\right)n\cos(n\phi_j)$$

$$M_{33}=m_{t_1}+m_{r_1}\tan^2\alpha_j-\mu\left(\cos\alpha_1+\sin\alpha_1\tan\alpha_1\right)(m_{r_1}-m_{t_1})\sin\alpha_1$$

$$M_{(j+2)(j+2)}=m_{t_j}+m_{r_j}\tan^2\alpha_j-\mu\left(\cos\alpha_j+\sin\alpha_j\tan\alpha_j\right)(m_{r_j}-m_{t_j})\sin\alpha_j$$

$$\mathrm{G}=\begin{bmatrix}
0 & -2M_{stat}n\Omega\left(n^2+1\right) & -2\rho_{b_1}S_{b_1}\frac{R_{stat}}{\pi}\Omega n\frac{\sin(n\phi_1)}{\cos\alpha_1} & \cdots\ \cdots\ \cdots & -2\rho_{b_j}S_{b_j}\frac{R_{stat}}{\pi}\Omega n\frac{\sin(n\phi_j)}{\cos\alpha_j}\\
2M_{stat}n\Omega\left(n^2+1\right) & 0 & 2\rho_{b_1}S_{b_1}\frac{R_{stat}}{\pi}\Omega n\frac{\cos(n\phi_1)}{\cos\alpha_1} & \cdots\ \cdots\ \cdots & 2\rho_{b_j}S_{b_j}\frac{R_{stat}}{\pi}\Omega n\frac{\cos(n\phi_j)}{\cos\alpha_j}\\
2\rho_{b_1}S_{b_1}\frac{R_{stat}}{\pi}\Omega n\frac{\sin(n\phi_1)}{\cos\alpha_1} & -2\rho_{b_1}S_{b_1}\frac{R_{stat}}{\pi}\Omega n\frac{\cos(n\phi_1)}{\cos\alpha_1} & 0 & \cdots\ \cdots\ \cdots & 0\\
\vdots & \vdots & \vdots & \ddots & \vdots\\
\vdots & \vdots & \vdots & \ddots & \vdots\\
\vdots & \vdots & \vdots & \ddots & \vdots\\
2\rho_{b_j}S_{b_j}\frac{R_{stat}}{\pi}\Omega n\frac{\sin(n\phi_j)}{\cos\alpha_j} & -2\rho_{b_j}S_{b_j}\frac{R_{stat}}{\pi}\Omega n\frac{\cos(n\phi_j)}{\cos\alpha_j} & 0 & \cdots\ \cdots\ \cdots & 0
\end{bmatrix}$$

$$\mathrm{R}=\begin{bmatrix}
R_{11} & R_{12} & R_{13} & \cdots & \cdots & \cdots & R_{1(j+2)}\\
R_{21} & R_{22} & R_{23} & \cdots & \cdots & \cdots & R_{2(j+2)}\\
R_{31} & R_{32} & 2\mu\rho_{b_1}S_{b_1}\frac{R_{stat}}{\pi}\Omega\left(\cos\alpha_1+\sin\alpha_1\tan\alpha_1\right)^2 & 0 & \cdots & 0 & 0\\
\vdots & \vdots & 0 & \ddots & 0 & \vdots & \vdots\\
\vdots & \vdots & \vdots & 0 & \ddots & 0 & \vdots\\
\vdots & \vdots & \vdots & \vdots & 0 & \ddots & \vdots\\
R_{(j+2)1} & R_{(j+2)2} & 0 & 0 & \cdots & 0 & 2\mu\rho_{b_j}S_{b_j}\frac{R_{stat}}{\pi}\Omega\left(\cos\alpha_j+\sin\alpha_j\tan\alpha_j\right)^2
\end{bmatrix}$$

$$R_{11}=-\sum_j 2\mu\rho_{b_j}S_{b_j}\frac{R_{stat}}{\pi}\Omega n\sin(n\phi_j)\tan\alpha_j\left(\left\{1+\frac{h}{2R_{stat}}\left(n^2-1\right)\right\}\cos(n\phi_j)-n\sin(n\phi_j)\tan\alpha_j\right)$$

$$R_{12}=\sum_j 2\mu\rho_{b_j}S_{b_j}\frac{R_{stat}}{\pi}\Omega n\cos(n\phi_j)\tan\alpha_j\left(\left\{1+\frac{h}{2R_{stat}}\left(n^2-1\right)\right\}\cos(n\phi_j)-n\sin(n\phi_j)\tan\alpha_j\right)$$

$$R_{21}=-\sum_j 2\mu\rho_{b_j}S_{b_j}\frac{R_{stat}}{\pi}\Omega n\sin(n\phi_j)\tan\alpha_j\left(\left\{1+\frac{h}{2R_{stat}}\left(n^2-1\right)\right\}\sin(n\phi_j)+n\cos(n\phi_j)\tan\alpha_j\right)$$

$$R_{22}=\sum_j 2\mu\rho_{b_j}S_{b_j}\frac{R_{stat}}{\pi}\Omega n\cos(n\phi_j)\tan\alpha_j\left(\left\{1+\frac{h}{2R_{stat}}\left(n^2-1\right)\right\}\sin(n\phi_j)+n\cos(n\phi_j)\tan\alpha_j\right)$$

$$R_{13}=-2\mu\rho_{b_1}S_{b_1}\frac{R_{stat}}{\pi}\Omega\left(\cos\alpha_1+\sin\alpha_1\tan\alpha_1\right)\left(\left\{1+\frac{h}{2R_{stat}}\left(n^2-1\right)\right\}\cos(n\phi_1)-n\sin(n\phi_1)\tan\alpha_1\right)$$

$$R_{1(j+2)}=-2\mu\rho_{b_j}S_{b_j}\frac{R_{stat}}{\pi}\Omega\left(\cos\alpha_j+\sin\alpha_j\tan\alpha_j\right)\left(\left\{1+\frac{h}{2R_{stat}}\left(n^2-1\right)\right\}\cos(n\phi_j)-n\sin(n\phi_j)\tan\alpha_j\right)$$

$$R_{23}=-2\mu\rho_{b_1}S_{b_1}\frac{R_{stat}}{\pi}\Omega\left(\cos\alpha_1+\sin\alpha_1\tan\alpha_1\right)\left(\left\{1+\frac{h}{2R_{stat}}\left(n^2-1\right)\right\}\sin(n\phi_1)+n\cos(n\phi_1)\tan\alpha_1\right)$$

$$R_{2(j+2)} = -2\mu\rho_{b_j} S_{b_j} \frac{R_{stat}}{\pi} \Omega\left(\cos\alpha_j + \sin\alpha_j \tan\alpha_j\right)\left(\left\{1+\frac{h}{2R_{stat}}\left(n^2-1\right)\right\}\sin(n\phi_j) + n\cos(n\phi_j)\tan\alpha_j\right)$$

$$R_{31} = 2\mu\rho_{b_1} S_{b_1} \frac{R_{stat}}{\pi} \Omega n \sin(n\phi_1)\tan\alpha_1\left(\cos\alpha_1 + \sin\alpha_1 \tan\alpha_1\right)$$

$$R_{32} = -2\mu\rho_{b_1} S_{b_1} \frac{R_{stat}}{\pi} \Omega n \cos(n\phi_1)\tan\alpha_1\left(\cos\alpha_1 + \sin\alpha_1 \tan\alpha_1\right)$$

$$R_{(j+2)1} = 2\mu\rho_{b_j} S_{b_j} \frac{R_{stat}}{\pi} \Omega n \sin(n\phi_j)\tan\alpha_j\left(\cos\alpha_j + \sin\alpha_j \tan\alpha_j\right)$$

$$R_{(j+2)2} = -2\mu\rho_{b_j} S_{b_j} \frac{R_{stat}}{\pi} \Omega n \cos(n\phi_j)\tan\alpha_j\left(\cos\alpha_j + \sin\alpha_j \tan\alpha_j\right)$$

$$\mathrm{K} = \begin{bmatrix} K_{11} & K_{12} & K_{13} & \cdots & \cdots & \cdots & K_{1(j+2)} \\ K_{21} & K_{22} & K_{23} & \cdots & \cdots & \cdots & K_{2(j+2)} \\ \left(k_{r_1} - m_{r_1}\Omega^2\right) n\sin(n\phi_1)\left(\frac{\tan\alpha_1}{\cos\alpha_1} - \mu\left(\cos\alpha_1 + \sin\alpha_1\tan\alpha_1\right)\right) & -\left(k_{r_1} - m_{r_1}\Omega^2\right) n\cos(n\phi_1)\left(\frac{\tan\alpha_1}{\cos\alpha_1} - \mu\left(\cos\alpha_1 + \sin\alpha_1\tan\alpha_1\right)\right) & K_{33} & 0 & \cdots & \cdots & 0 \\ \vdots & \vdots & 0 & \ddots & 0 & \cdots & 0 \\ \vdots & \vdots & \vdots & 0 & \ddots & \ddots & \vdots \\ \vdots & \vdots & \vdots & \vdots & \ddots & \ddots & 0 \\ \left(k_{r_j} - m_{r_j}\Omega^2\right) n\sin(n\phi_j)\left(\frac{\tan\alpha_j}{\cos\alpha_j} - \mu\left(\cos\alpha_j + \sin\alpha_j\tan\alpha_j\right)\right) & -\left(k_{r_j} - m_{r_j}\Omega^2\right) n\cos(n\phi_j)\left(\frac{\tan\alpha_j}{\cos\alpha_j} - \mu\left(\cos\alpha_j + \sin\alpha_j\tan\alpha_j\right)\right) & 0 & 0 & \cdots & 0 & K_{(j+2)(j+2)} \end{bmatrix}$$

$$K_{11} = K_{stat} n^2 (n^2-1)^2 - M_{stat} n^2 \Omega^2\left(n^2+1\right) + \sum_j (k_{r_j} - m_{r_j}\Omega^2) n^2 \frac{\sin^2(n\phi_j)}{\cos^2\alpha_j} + \mu\sum_j\left(k_{r_j} - m_{r_j}\Omega^2\right) n\sin(n\phi_j)\left(\left\{1+\frac{h}{2R_{stat}}\left(n^2-1\right)\right\}\cos(n\phi_j) - n\sin(n\phi_j)\tan\alpha_j\right)$$

$$K_{12} = -\sum_j (k_{r_j} - m_{r_j}\Omega^2) n^2 \frac{\sin(n\phi_j)\cos(n\phi_j)}{\cos^2\alpha_j} - \mu\sum_j\left(k_{r_j} - m_{r_j}\Omega^2\right) n\cos(n\phi_j)\left(\left\{1+\frac{h}{2R_{stat}}\left(n^2-1\right)\right\}\cos(n\phi_j) - n\sin(n\phi_j)\tan\alpha_j\right)$$

$$K_{21} = -\sum_j (k_{r_j} - m_{r_j}\Omega^2) n^2 \frac{\sin(n\phi_j)\cos(n\phi_j)}{\cos^2\alpha_j} + \mu\sum_j\left(k_{r_j} - m_{r_j}\Omega^2\right) n\sin(n\phi_j)\left(\left\{1+\frac{h}{2R_{stat}}\left(n^2-1\right)\right\}\sin(n\phi_j) + n\cos(n\phi_j)\tan\alpha_j\right)$$

$$K_{22} = K_{stat} n^2 (n^2-1)^2 - M_{stat} n^2 \Omega^2\left(n^2+1\right) + \sum_j (k_{r_j} - m_{r_j}\Omega^2) n^2 \frac{\cos^2(n\phi_j)}{\cos^2\alpha_j} - \mu\sum_j\left(k_{r_j} - m_{r_j}\Omega^2\right) n\cos(n\phi_j)\left(\left\{1+\frac{h}{2R_{stat}}\left(n^2-1\right)\right\}\sin(n\phi_j) + n\cos(n\phi_j)\tan\alpha_j\right)$$

$$K_{13} = \left(k_{r_1} - m_{r_1}\Omega^2\right) n\sin(n\phi_1)\frac{\tan\alpha_1}{\cos\alpha_1} + \mu\left(\left\{1+\frac{h}{2R_{stat}}\left(n^2-1\right)\right\}\cos(n\phi_1) - n\sin(n\phi_1)\tan\alpha_1\right)\left[\left(k_{r_1} - m_{r_1}\Omega^2\right) - \left(k_{t_1} - \left(m_{t_1} - \rho_{b_1} I_{b_1}\frac{\pi^2}{8R_{stat}}\right)\Omega^2\right)\right]\sin\alpha_1$$

$$K_{1(j+2)} = \left(k_{r_j} - m_{r_j}\Omega^2\right) n\sin(n\phi_j)\frac{\tan\alpha_j}{\cos\alpha_j} + \mu\left(\left\{1+\frac{h}{2R_{stat}}\left(n^2-1\right)\right\}\cos(n\phi_j) - n\sin(n\phi_j)\tan\alpha_j\right)\left[\left(k_{r_j} - m_{r_j}\Omega^2\right) - \left(k_{t_j} - \left(m_{t_j} - \rho_{b_j} I_{b_j}\frac{\pi^2}{8R_{stat}}\right)\Omega^2\right)\right]\sin\alpha_j$$

$$K_{23} = -\left(k_{r_1} - m_{r_1}\Omega^2\right) n\cos(n\phi_1)\frac{\tan\alpha_1}{\cos\alpha_1} + \mu\left(\left\{1+\frac{h}{2R_{stat}}\left(n^2-1\right)\right\}\sin(n\phi_1) + n\cos(n\phi_1)\tan\alpha_1\right)\left[\left(k_{r_1} - m_{r_1}\Omega^2\right) - \left(k_{t_1} - \left(m_{t_1} - \rho_{b_1} I_{b_1}\frac{\pi^2}{8R_{stat}}\right)\Omega^2\right)\right]\sin\alpha_1$$

$$K_{2(j+2)} = -\left(k_{r_j} - m_{r_j}\Omega^2\right) n\cos(n\phi_j)\frac{\tan\alpha_j}{\cos\alpha_j} + \mu\left(\left\{1+\frac{h}{2R_{stat}}\left(n^2-1\right)\right\}\sin(n\phi_j) + n\cos(n\phi_j)\tan\alpha_j\right)\left[\left(k_{r_j} - m_{r_j}\Omega^2\right) - \left(k_{t_j} - \left(m_{t_j} - \rho_{b_j} I_{b_j}\frac{\pi^2}{8R_{stat}}\right)\Omega^2\right)\right]\sin\alpha_j$$

$$K_{33} = \left[k_{r_1}\tan^2\alpha_1 + k_{t_1}\right] - \left[m_{r_1}\tan^2\alpha_1 + \left(m_{t_1} - \rho_{b_1} I_{b_1}\frac{\pi^2}{8R_{stat}}\right)\right]\Omega^2 - \mu\left(\cos\alpha_1 + \sin\alpha_1\tan\alpha_1\right)\left[\left(k_{r_1} - m_{r_1}\Omega^2\right) - \left(k_{t_1} - \left(m_{t_1} - \rho_{b_1} I_{b_1}\frac{\pi^2}{8R_{stat}}\right)\Omega^2\right)\right]\sin\alpha_1$$

$$K_{(j+2)(j+2)} = \left[k_{r_j}\tan^2\alpha_j + k_{t_j}\right] - \left[m_{r_j}\tan^2\alpha_j + \left(m_{t_j} - \rho_{b_j} I_{b_j}\frac{\pi^2}{8R_{stat}}\right)\right]\Omega^2 - \mu\left(\cos\alpha_j + \sin\alpha_j\tan\alpha_j\right)\left[\left(k_{r_j} - m_{r_j}\Omega^2\right) - \left(k_{t_j} - \left(m_{t_j} - \rho_{b_j} I_{b_j}\frac{\pi^2}{8R_{stat}}\right)\Omega^2\right)\right]\sin\alpha_j$$

$$
\mathrm{F} = \left\{ \begin{array}{c}
\sum_j \left( 4\rho_{b_j} S_{b_j} \frac{R_{stat}^2}{\pi^2}\Omega^2 + N_U \right) n \frac{\sin(n\phi_j)}{\cos\alpha_j} + \sum_j \mu \left[ \left\{ 1 + \frac{h}{2R_{stat}}\left(n^2-1\right) \right\} \cos(n\phi_j) - n\sin(n\phi_j)\tan\alpha_j \right] \left( 4\rho_{b_j} S_{b_j} \frac{R_{stat}^2}{\pi^2}\Omega^2 + N_U \right) \\
-\sum_j \left( 4\rho_{b_j} S_{b_j} \frac{R_{stat}^2}{\pi^2}\Omega^2 + N_U \right) n \frac{\cos(n\phi_j)}{\cos\alpha_j} + \sum_j \mu \left[ \left\{ 1 + \frac{h}{2R_{stat}}\left(n^2-1\right) \right\} \sin(n\phi_j) + n\cos(n\phi_j)\tan\alpha_j \right] \left( 4\rho_{b_j} S_{b_j} \frac{R_{stat}^2}{\pi^2}\Omega^2 + N_U \right) \\
\left( 4\rho_{b_1} S_{b_1} \frac{R_{stat}^2}{\pi^2}\Omega^2 + N_U \right) \tan\alpha_1 - \mu\left[\cos\alpha_1 + \sin\alpha_1 \tan\alpha_1\right] \left( 4\rho_{b_1} S_{b_1} \frac{R_{stat}^2}{\pi^2}\Omega^2 + N_U \right) \\
\vdots \\
\vdots \\
\vdots \\
\left( 4\rho_{b_j} S_{b_j} \frac{R_{stat}^2}{\pi^2}\Omega^2 + N_U \right) \tan\alpha_j - \mu\left[\cos\alpha_j + \sin\alpha_j \tan\alpha_j\right] \left( 4\rho_{b_j} S_{b_j} \frac{R_{stat}^2}{\pi^2}\Omega^2 + N_U \right)
\end{array} \right\}
$$

In these expressions, $m_{r_j} = \rho_{b_j} S_{b_j} \frac{R_{stat}}{2}$ and $m_{t_j} = \rho_{b_j} S_{b_j} R_{stat}\left(\frac{3}{2} - \frac{4}{\pi}\right) + \rho_{b_j} I_{b_j} \frac{\pi^2}{8R_{stat}}$ are the modal mass of traction/compression and flexure, respectively. $k_{r_j} = E_{b_j} S_{b_j} \frac{\pi^2}{8R_{stat}}$ and $k_{t_j} = E_{b_j} I_{b_j} \frac{\pi^4}{32R_{stat}^3}$ are the modal stiffness of traction/compression and flexure, respectively.

## APPENDIX B

Expressions of the kinetic energy and potential energy, as well as of the work of the rubbing strength associated with the simplified model of rotating spring-masses featuring two degrees of freedom rubbing on the flexible inextensible ring.

The expression of system kinetic energy is given by:

$$T=\frac{1}{2}\int_{-\Omega t}^{2\pi-\Omega t}\rho_{stat}S_{stat}\left\{\left[\dot{u}_s(\phi,t)-\Omega\frac{\partial u_s}{\partial\phi}(\phi,t)\right]^2+\left[\dot{w}(\phi,t)-\Omega\frac{\partial w}{\partial\phi}(\phi,t)\right]^2\right\}R_{stat}d\phi+\sum_j\frac{1}{2}m_j\left[\dot{u}_s^2(\phi,t)+\Omega^2\left(R_{stat}-u_s(\phi,t)\right)^2\right]\delta\left(\phi-\phi_j\right)$$

$$+\sum_j\frac{1}{2}m_j\left[\dot{x}^2(\phi,t)+\Omega^2x^2\right]\delta\left(\phi-\phi_j\right)+\sum_j m_j\Omega\left[\dot{u}_s(\phi,t)x(\phi,t)+\dot{x}(\phi,t)\left(R_{stat}-u_s(\phi,t)\right)\right]\delta\left(\phi-\phi_j\right)$$

The expression of system potential energy is given by:

$$\gamma=\frac{1}{2}\int_{-\Omega t}^{2\pi-\Omega t}\frac{E_{stat}I_{stat}}{R_{stat}^3}\left\{\frac{\partial^2u_s}{\partial\phi^2}(\phi,t)+u_s(\phi,t)\right\}^2d\phi+\sum_j\frac{1}{2}k_{r_j}u_s^2(\phi,t)\delta\left(\phi-\phi_j\right)+\sum_j\frac{1}{2}k_{t_j}x_j^2(\phi,t)\delta\left(\phi-\phi_j\right)$$

The expression of the rubbing work can now be given by:

$$W_{ext}=\sum_j T_{m_j\to stat}\left[\omega(\phi,t)\left\{1-\frac{h}{2R_{stat}}\right\}-\frac{h}{2R_{stat}}\frac{\partial u_s(\phi,t)}{\partial\phi}-x_j(\phi,t)\right]\delta(\phi-\phi_j)$$

with: $T_{m_j\to stat}=\mu\left[N_U+m_j\left(\ddot{u}_s(\phi,t)-\Omega^2u_s(\phi,t)+\Omega^2R_{stat}+2\dot{x}_j(\phi,t)\Omega\right)+k_{r_j}u_s(\phi,t)\right]\delta(\phi-\phi_j)$ being the rubbing strength of mass $m_j$ on the stator.

The same remarks as those offered in Appendix A can be forwarded here concerning validity conditions of the rubbing model. Here again, only one mode shape of the stator has been considered at a time.
The matrix equation of system dynamic behaviour is as follows:

$$\mathbf{M}\ddot{\mathbf{X}}+(\mathbf{G}+\mathbf{R})\dot{\mathbf{X}}+\mathbf{KX}=\mathbf{F}.$$

with: $\mathbf{X}^T=\left\{A_n\quad B_n\quad x_1\quad\cdots\quad\cdots\quad\cdots\quad x_j\right\}$

$$\mathrm{M}=\begin{bmatrix} M_{11} & M_{12} & 0 & \cdots & \cdots & \cdots & 0\\ M_{21} & M_{22} & 0 & \cdots & \cdots & \cdots & 0\\ -\mu m_1 n\sin(n\phi_1) & \mu m_1 n\cos(n\phi_1) & m_1 & 0 & \cdots & \cdots & 0\\ \vdots & \vdots & 0 & \ddots & 0 & \cdots & 0\\ \vdots & \vdots & \vdots & 0 & \ddots & \ddots & \vdots\\ \vdots & \vdots & \vdots & \vdots & \ddots & \ddots & 0\\ -\mu m_j n\sin(n\phi_j) & \mu m_j n\cos(n\phi_j) & 0 & 0 & \cdots & 0 & m_j \end{bmatrix}$$

$$M_{11}=M_{stat}\left(n^2+1\right)+\sum_j m_jn^2\sin^2(n\phi_j)+\mu\left\{1+\frac{h}{2R_{stat}}\left(n^2-1\right)\right\}\sum_j m_jn\sin(n\phi_j)\cos(n\phi_j)$$

$$M_{12}=-\sum_j m_jn^2\sin(n\phi_j)\cos(n\phi_j)-\mu\left\{1+\frac{h}{2R_{stat}}\left(n^2-1\right)\right\}\sum_j m_jn\cos^2(n\phi_j)$$

$$M_{21}=-\sum_j m_jn^2\sin(n\phi_j)\cos(n\phi_j)+\mu\left\{1+\frac{h}{2R_{stat}}\left(n^2-1\right)\right\}\sum_j m_jn\sin^2(n\phi_j)$$

$$M_{22}=M_{stat}\left(n^2+1\right)+\sum_j m_jn^2\cos^2(n\phi_j)-\mu\left\{1+\frac{h}{2R_{stat}}\left(n^2-1\right)\right\}\sum_j m_jn\sin(n\phi_j)\cos(n\phi_j)$$

$$
\mathrm{G}=\begin{bmatrix}
0 & -2M_{stat}n\Omega\left(n^2+1\right) & -2m_1\Omega n\sin(n\phi_1) & \cdots\ \cdots\ \cdots & -2m_j\Omega n\sin(n\phi_j)\\
2M_{stat}n\Omega\left(n^2+1\right) & 0 & 2m_1\Omega n\cos(n\phi_1) & \cdots\ \cdots\ \cdots & 2m_j\Omega n\cos(n\phi_j)\\
2m_1\Omega n\sin(n\phi_1) & -2m_1\Omega n\cos(n\phi_1) & 0 & \cdots\ \cdots\ \cdots & 0\\
\vdots & \vdots & \vdots & \ddots & \vdots\\
\vdots & \vdots & \vdots & \ddots & \vdots\\
\vdots & \vdots & \vdots & \ddots & \vdots\\
2m_j\Omega n\sin(n\phi_j) & -2m_j\Omega n\cos(n\phi_j) & 0 & \cdots\ \cdots\ \cdots & 0
\end{bmatrix}
$$

$$
\mathrm{R}=\begin{bmatrix}
0 & 0 & -2\mu m_1\Omega\cos(n\phi_1)\left\{1+\dfrac{h}{2R_{stat}}\left(n^2-1\right)\right\} & \cdots & \cdots & \cdots & -2\mu m_j\Omega\cos(n\phi_j)\left\{1+\dfrac{h}{2R_{stat}}\left(n^2-1\right)\right\}\\
\vdots & \vdots & -2\mu m_1\Omega\sin(n\phi_1)\left\{1+\dfrac{h}{2R_{stat}}\left(n^2-1\right)\right\} & \cdots & \cdots & \cdots & -2\mu m_j\Omega\sin(n\phi_j)\left\{1+\dfrac{h}{2R_{stat}}\left(n^2-1\right)\right\}\\
\vdots & \vdots & 2\mu m_1\Omega & 0 & \cdots & 0 & 0\\
\vdots & \vdots & 0 & \ddots & 0 & \vdots & \vdots\\
\vdots & \vdots & \vdots & 0 & \ddots & 0 & \vdots\\
\vdots & \vdots & \vdots & \vdots & 0 & \ddots & \vdots\\
0 & 0 & 0 & 0 & \cdots & 0 & 2\mu m_j\Omega
\end{bmatrix}
$$

$$
\mathrm{K}=\begin{bmatrix}
K_{11} & K_{12} & 0 & \cdots & \cdots & \cdots & 0\\
K_{21} & K_{22} & 0 & \cdots & \cdots & \cdots & 0\\
-\mu\left(k_{r_1}-m_1\Omega^2\right)n\sin(n\phi_1) & \mu\left(k_{r_1}-m_1\Omega^2\right)n\cos(n\phi_1) & k_{t_1}-m_1\Omega^2 & 0 & \cdots & \cdots & 0\\
\vdots & \vdots & 0 & \ddots & \ddots & \cdots & 0\\
\vdots & \vdots & \vdots & \ddots & \ddots & \ddots & \vdots\\
\vdots & \vdots & \vdots & \vdots & \ddots & \ddots & 0\\
-\mu\left(k_{r_j}-m_j\Omega^2\right)n\sin(n\phi_j) & \mu\left(k_{r_j}-m_j\Omega^2\right)n\cos(n\phi_j) & 0 & 0 & \cdots & 0 & k_{t_j}-m_j\Omega^2
\end{bmatrix}
$$

$$
K_{11}=K_{stat}n^2(n^2-1)^2-M_{stat}n^2\Omega^2\left(n^2+1\right)+\sum_j(k_{r_j}-m_j\Omega^2)n^2\sin^2(n\phi_j)+\mu\left\{1+\frac{h}{2R_{stat}}\left(n^2-1\right)\right\}\sum_j\left(k_{r_j}-m_j\Omega^2\right)n\sin(n\phi_j)\cos(n\phi_j)
$$

$$
K_{12}=-\sum_j(k_{r_j}-m_j\Omega^2)n^2\sin(n\phi_j)\cos(n\phi_j)-\mu\left\{1+\frac{h}{2R_{stat}}\left(n^2-1\right)\right\}\sum_j\left(k_{r_j}-m_j\Omega^2\right)n\cos^2(n\phi_j)
$$

$$
K_{21}=-\sum_j(k_{r_j}-m_j\Omega^2)n^2\sin(n\phi_j)\cos(n\phi_j)+\mu\left\{1+\frac{h}{2R_{stat}}\left(n^2-1\right)\right\}\sum_j\left(k_{r_j}-m_j\Omega^2\right)n\sin^2(n\phi_j)
$$

$$
K_{22}=K_{stat}n^2(n^2-1)^2-M_{stat}n^2\Omega^2\left(n^2+1\right)+\sum_j(k_{r_j}-m_j\Omega^2)n^2\cos^2(n\phi_j)-\mu\left\{1+\frac{h}{2R_{stat}}\left(n^2-1\right)\right\}\sum_j\left(k_{r_j}-m_j\Omega^2\right)n\sin(n\phi_j)\cos(n\phi_j)
$$

$$
\mathrm{F}=\begin{Bmatrix}
\sum_j\left(m_jR_{stat}\Omega^2+N_U\right)n\sin(n\phi_j)+\mu\left\{1+\dfrac{h}{2R_{stat}}\left(n^2-1\right)\right\}\sum_j\left(m_j\Omega^2R_{stat}+N_U\right)\cos(n\phi_j)\\
-\sum_j\left(m_jR_{stat}\Omega^2+N_U\right)n\cos(n\phi_j)+\mu\left\{1+\dfrac{h}{2R_{stat}}\left(n^2-1\right)\right\}\sum_j\left(m_j\Omega^2R_{stat}+N_U\right)\sin(n\phi_j)\\
-\mu\left(m_1\Omega^2R_{stat}+N_U\right)\\
\vdots\\
\vdots\\
\vdots\\
-\mu\left(m_j\Omega^2R_{stat}+N_U\right)
\end{Bmatrix}
$$

Under such conditions, differences between this system and the beam model stem from spin-softening terms since those associated with the beam model do not take into account the entire flexural modal mass, but instead $\left(m_{t_j}-\rho_{b_j}I_{b_j}\dfrac{\pi^2}{8R_{stat}}\right)$. Another difference concerning both the matrix $\mathbf{R}$ and gyroscopic terms has also been identified.

| Angular position of second blade | $\Omega_{X_1}(RPM)$ | $\Omega_{X_2}(RPM)$ |
|---|---|---|
| 87° | 1145 | 1266 |
| 90° | 1060 | 1139 |
| 110° | 955 | 973 |
| 120° | 1060 | 1139 |
| 123° | 1145 | 1266 |

Table 1 Coupling rotational speeds between two spring-masses with $m = 100kg$ and $k_r = k_t = 1.10^6 N.m^{-1}$, the first one being at 60° in the rotating frame and the other one, in the first coupling angular region (87° - 123°)

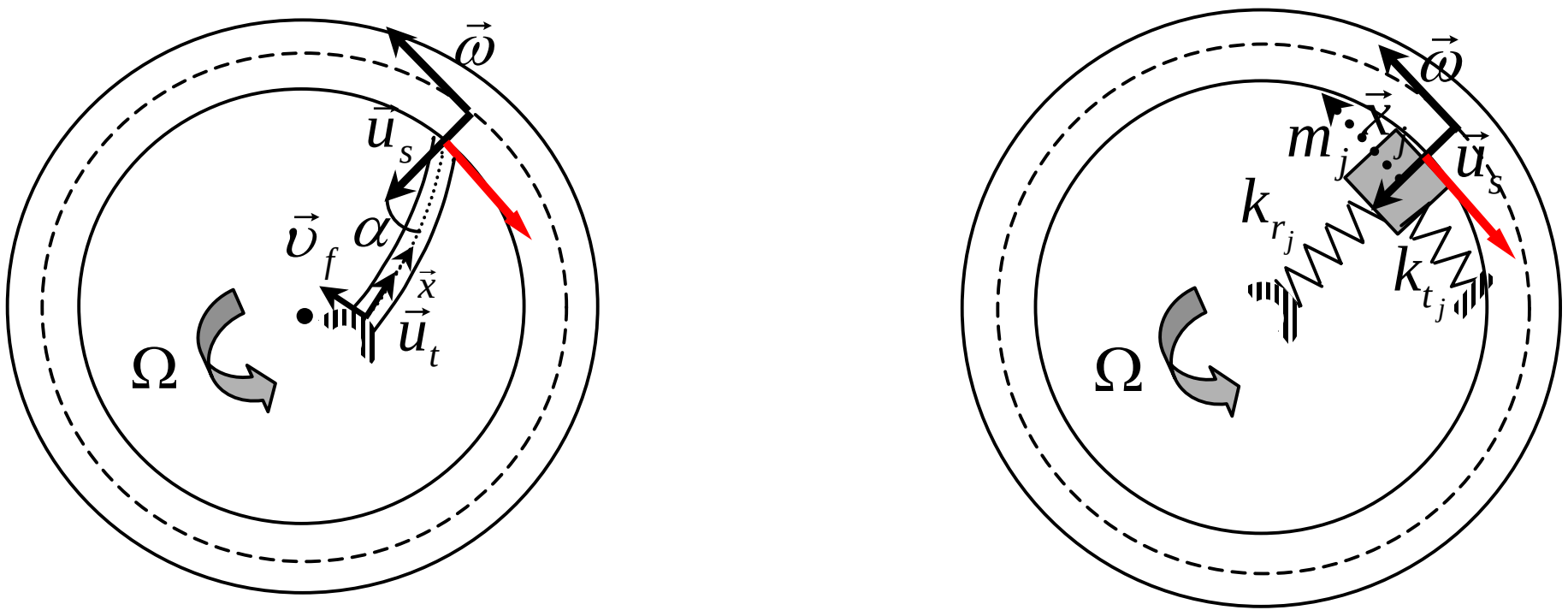

(a) (b)

Figure 1 a) Model of Euler-Bernoulli beam rubbing on an elastic ring, b) model of ring rubbed by one rotating load having two degrees of freedom

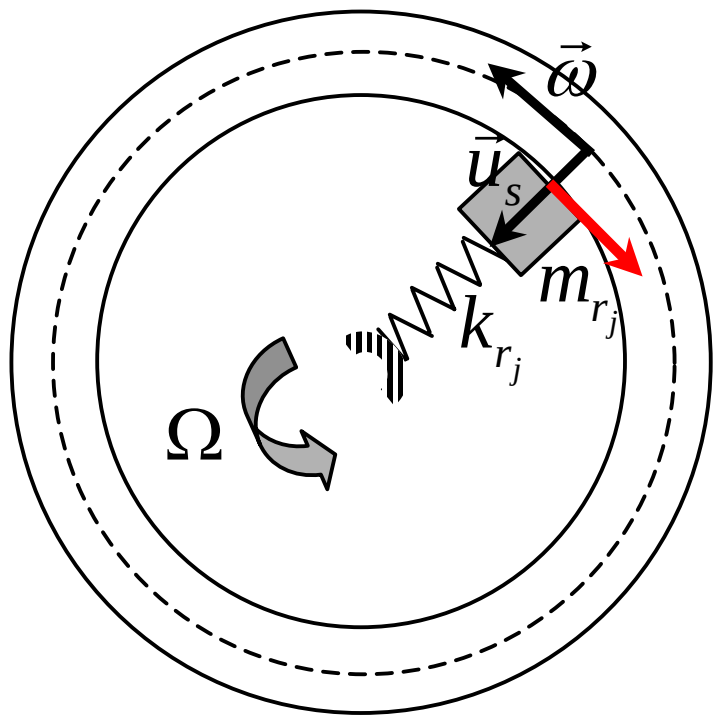

Figure 2 Model of radial spring-mass rubbing against a ring

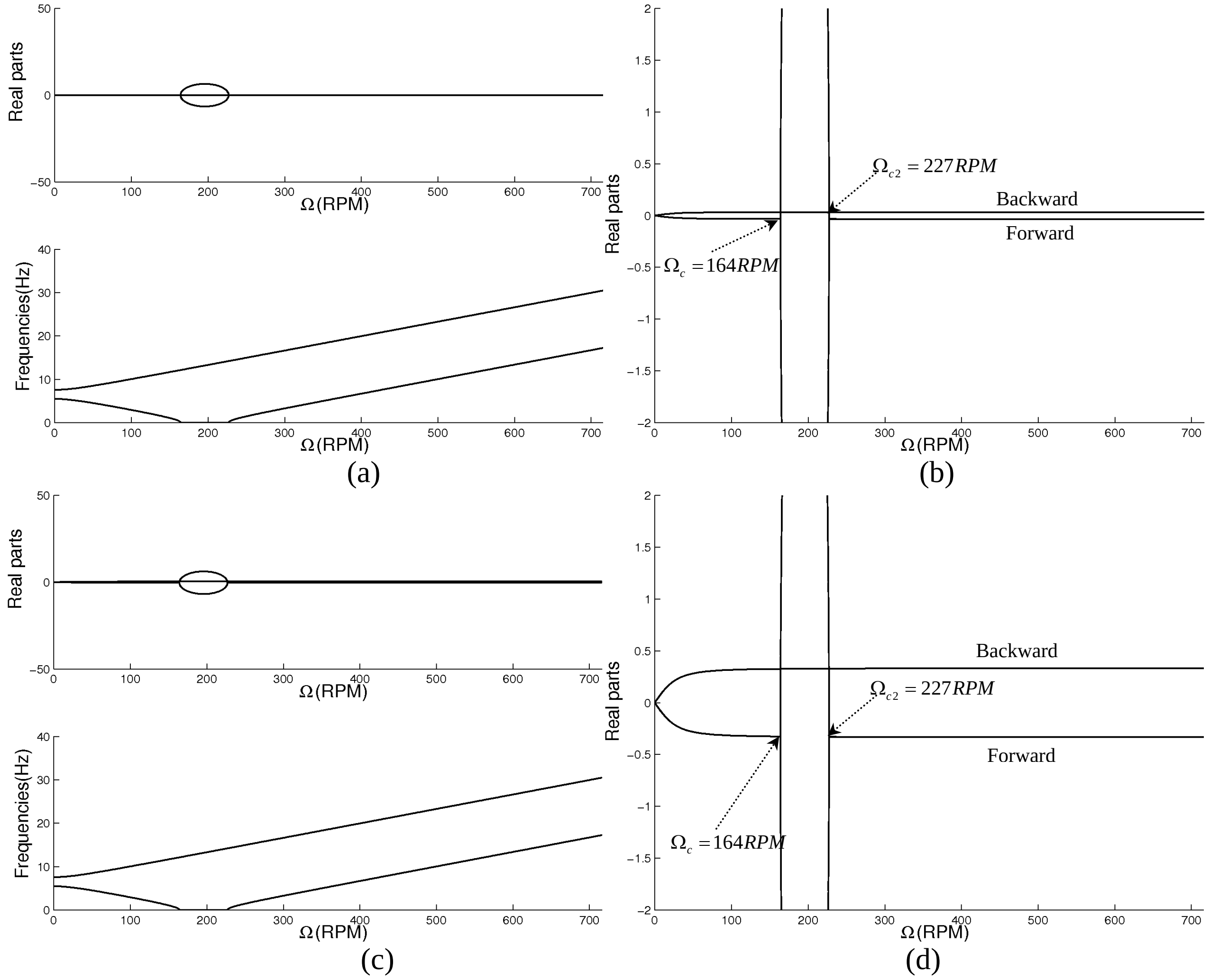

Figure 3 Stability analysis for a radial stiffness of $k_r = 1.10^6\,N.m^{-1}$ rubbing on the two nodal diameter mode shape of the ring with a) $\mu = 0.01$ , b) being the associated zoom and c) $\mu = 0.1$, d) being the associated zoom

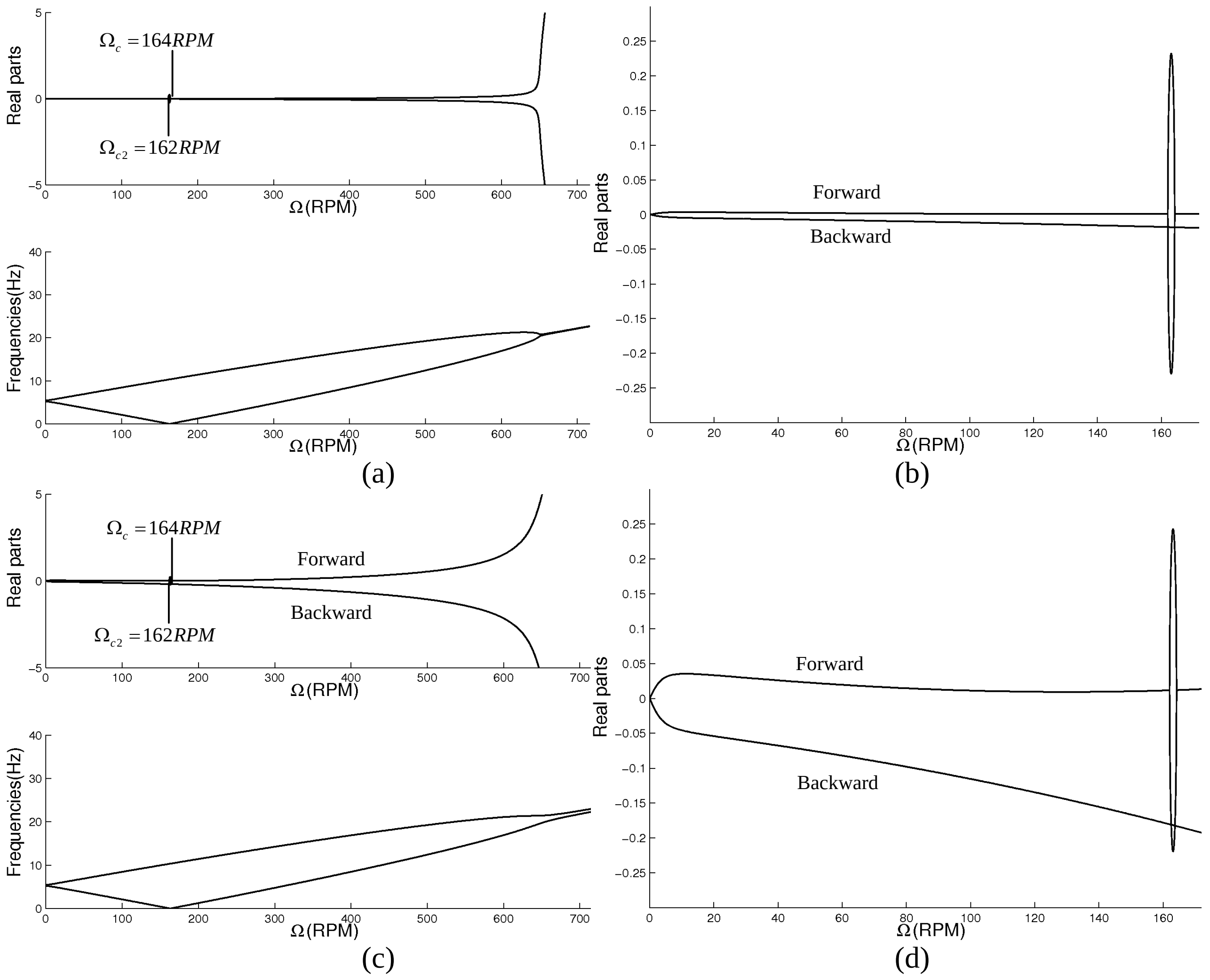

Figure 4 Stability analysis for a mass of 100kg rubbing against the two nodal diameter mode shape of the ring with a) $\mu = 0.01$ , b) being the associated zoom and c) $\mu = 0.1$, d) being the associated zoom

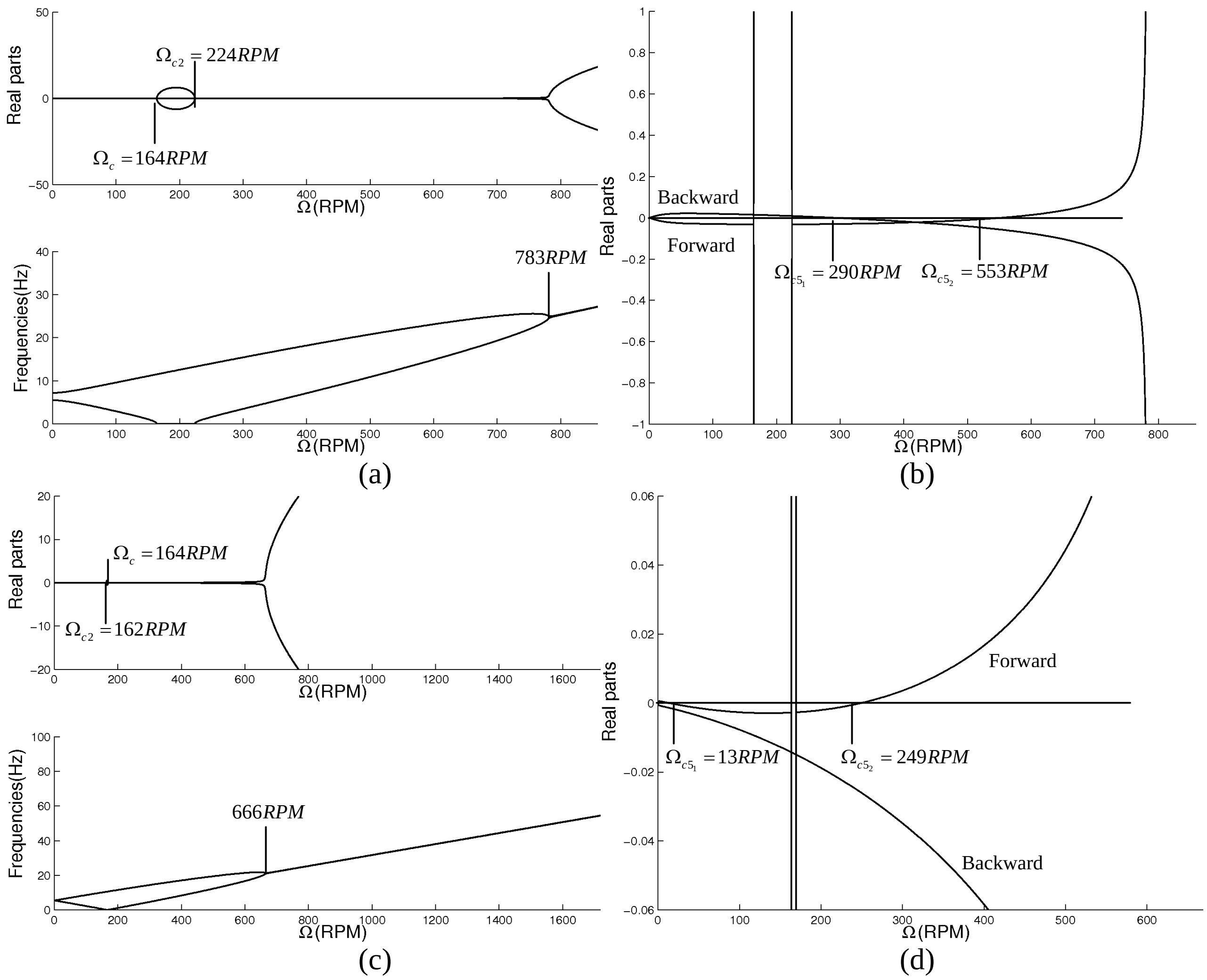

Figure 5 Stability analysis of the two nodal diameter mode shape of the ring excited by a rubbing ($\mu = 0.01$) radial spring-mass with a) $\omega_r = 100 rad/s$, b) being the associated zoom and c) $\omega_r = 31.6 rad/s$, d) being the associated zoom whereas $\omega_n = 34.4 rad/s$

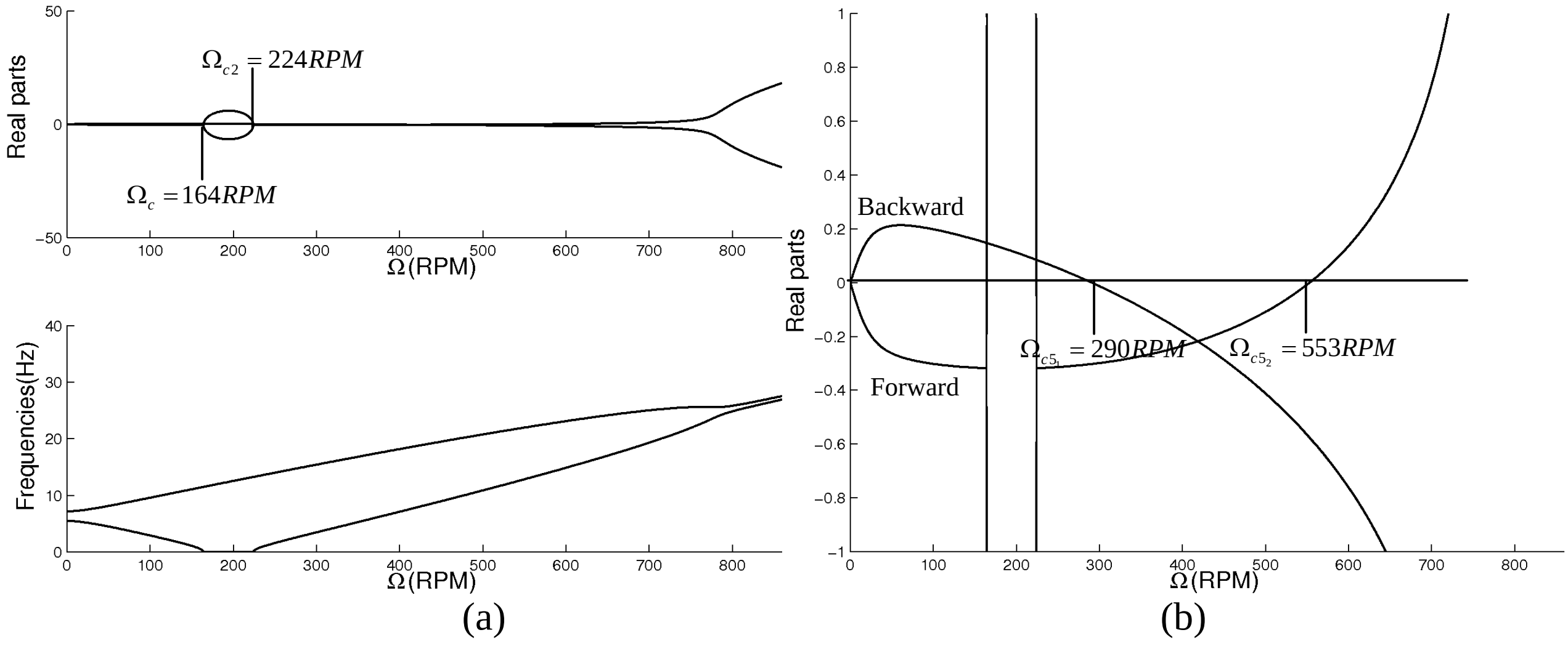

Figure 6 a) Stability analysis of the two nodal diameter mode shape of the ring excited by a rubbing radial spring-mass with $\omega_r = 100 rad/s$ and $\mu = 0.1$, b) being the associated zoom

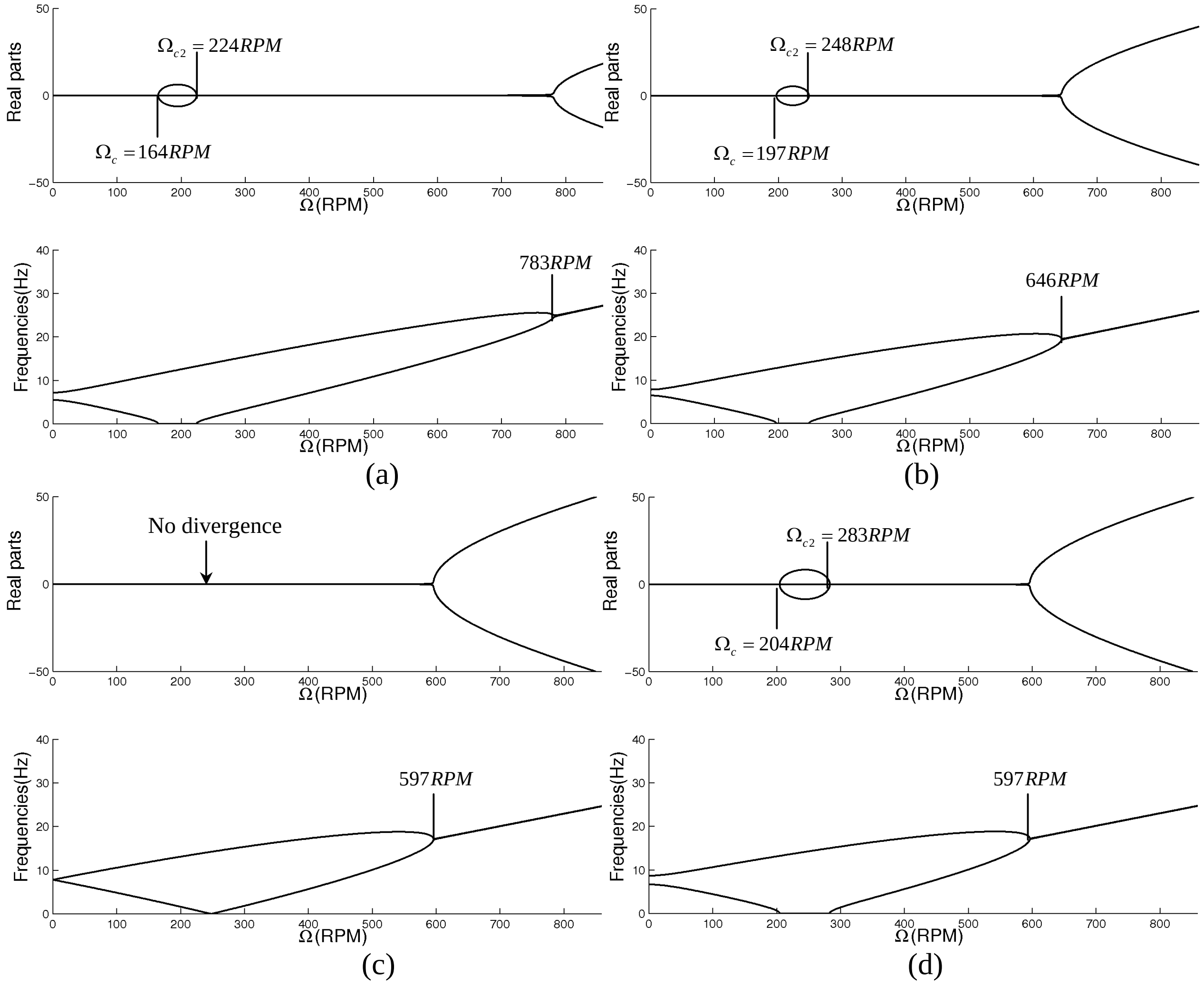

Figure 7 Stability analysis of the two nodal diameter mode shape of the ring excited by a) one radial spring-mass b) two radial spring-masses separated from 60° from each other, c) three radial spring-masses separated from 60° from each other, d) three radial spring-masses two being at 60° form each other and the third one being at 180° from one of the latter two

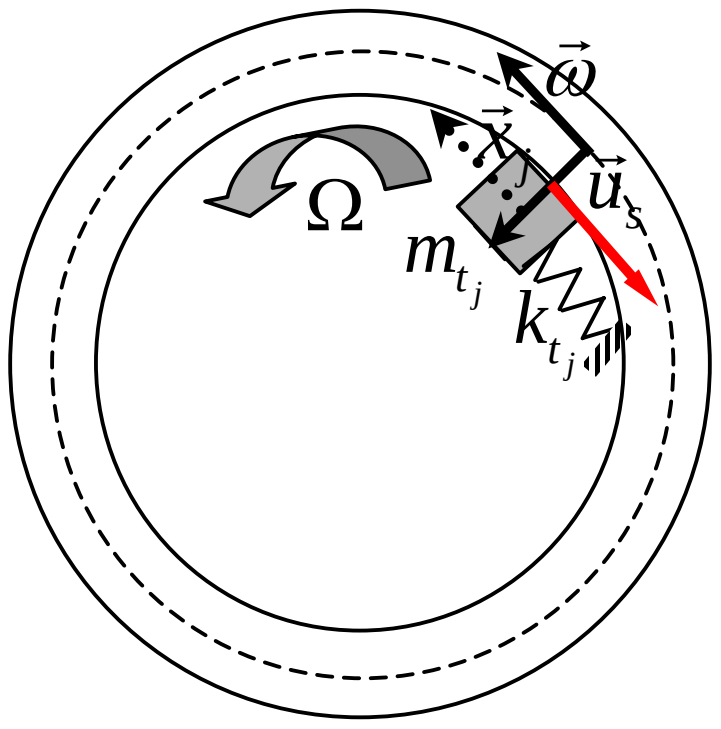

Figure 8 Model of rubbing rotating spring-mass tangent to the ring

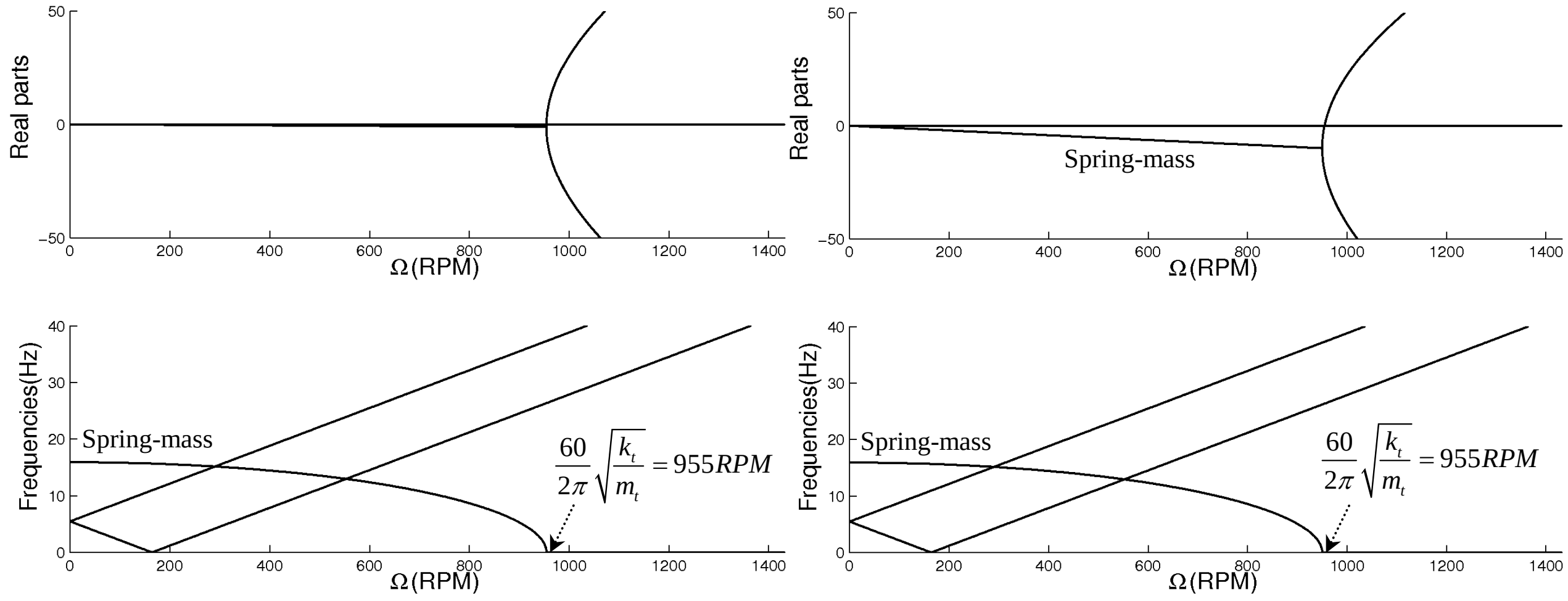

(a) (b)

Figure 9 Stability analysis for a tangent spring-mass rubbing against the two nodal diameter mode shape of the ring with a) $\mu = 0.01$, and b) $\mu = 0.1$

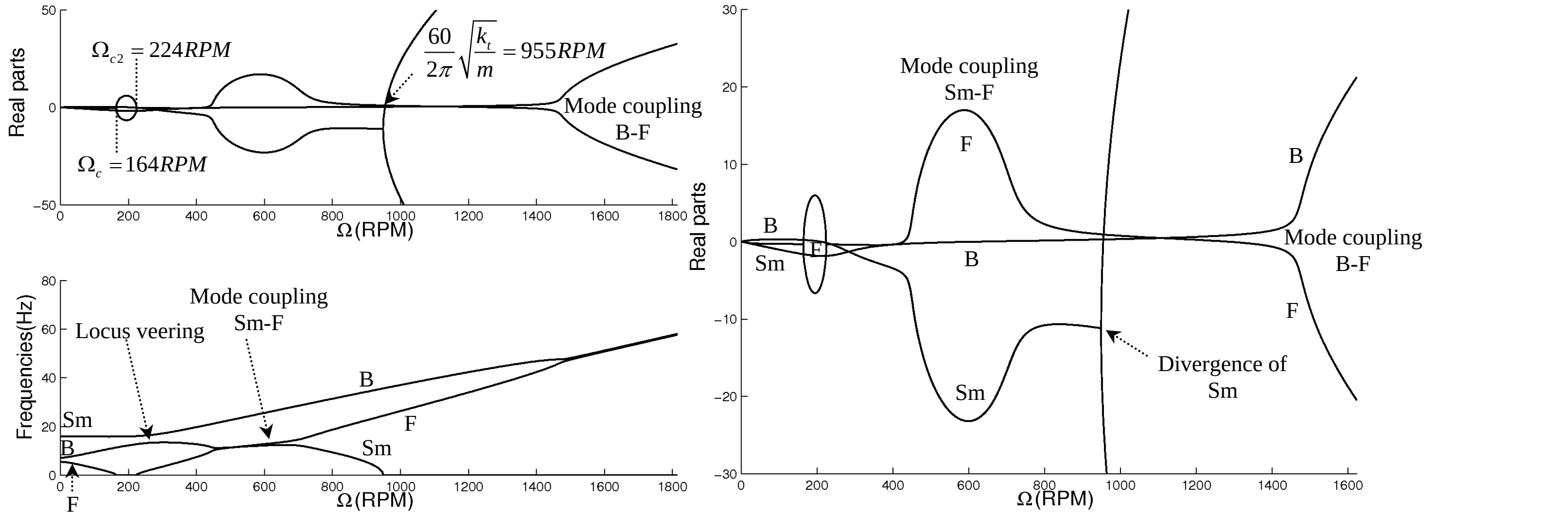

**(a)** **(b)**

Figure 10 a) Stability analysis of the two nodal diameter mode shape of the ring rubbed by a spring-mass having $m = 100kg$, $k_r = k_t = 1.10^6\,N.m^{-1}$ and $\mu = 0.1$, b) being the associated zoom

Sm = Spring-mass, F = Forward, B = Backward

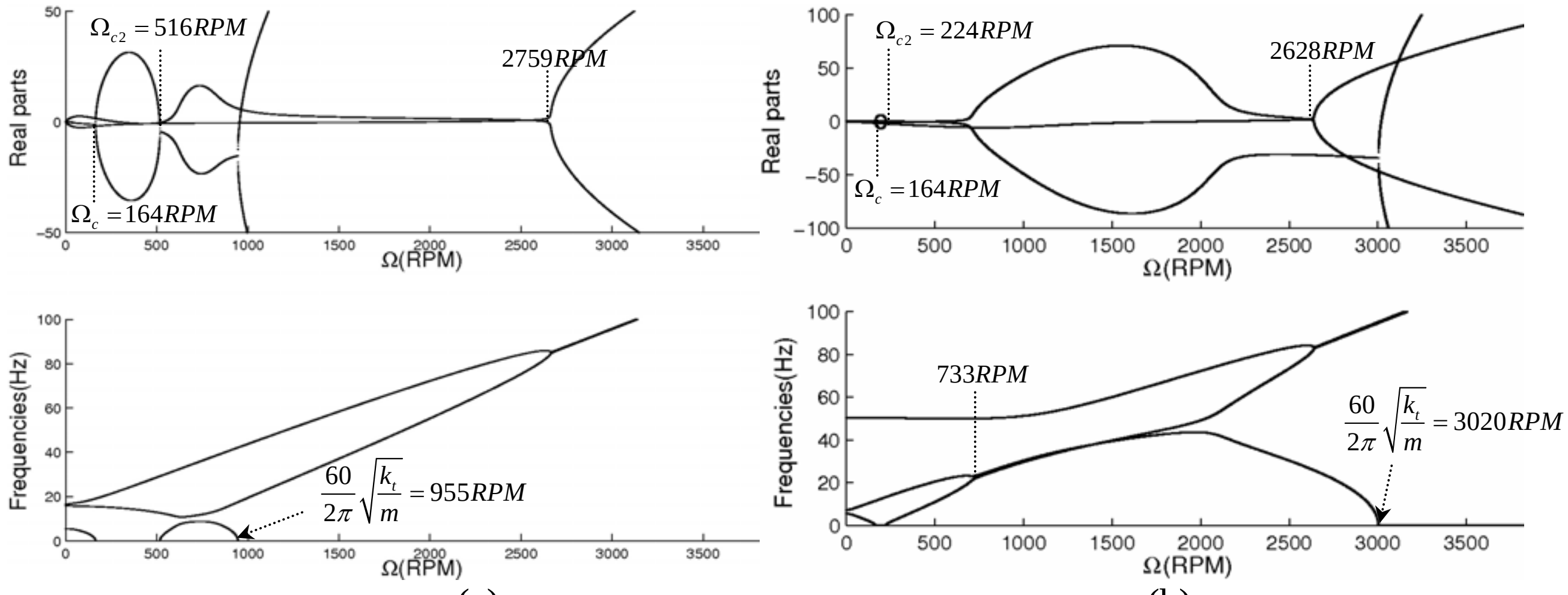

(a) (b)

Figure 11 Stability analysis of the two nodal diameter mode shape of the ring rubbed by a spring-mass of $m = 100kg$, a) $k_r = 1.10^7\,N.m^{-1}$, $k_t = 1.10^6\,N.m^{-1}$ b) $k_r = 1.10^6\,N.m^{-1}$, $k_t = 1.10^7\,N.m^{-1}$ and $\mu = 0.1$

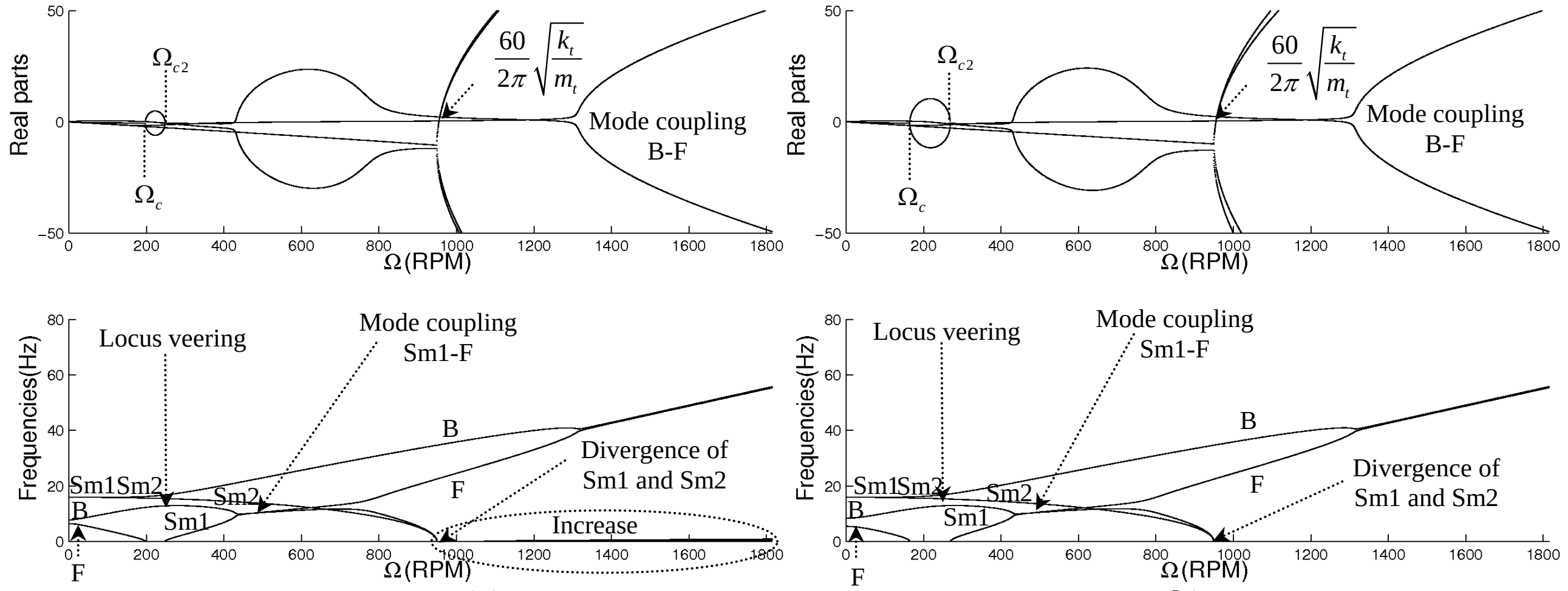

(a) (b)

Figure 12 Stability analysis of the two nodal diameter mode shape of the ring rubbed by two spring-masses having $m = 100kg$ , $k_r = k_t = 1.10^6\, N.m^{-1}$ and $\mu = 0.1$, a) separated from 60° from each other b) separated from 180° from each other

Sm = Spring-mass, F = Forward, B = Backward

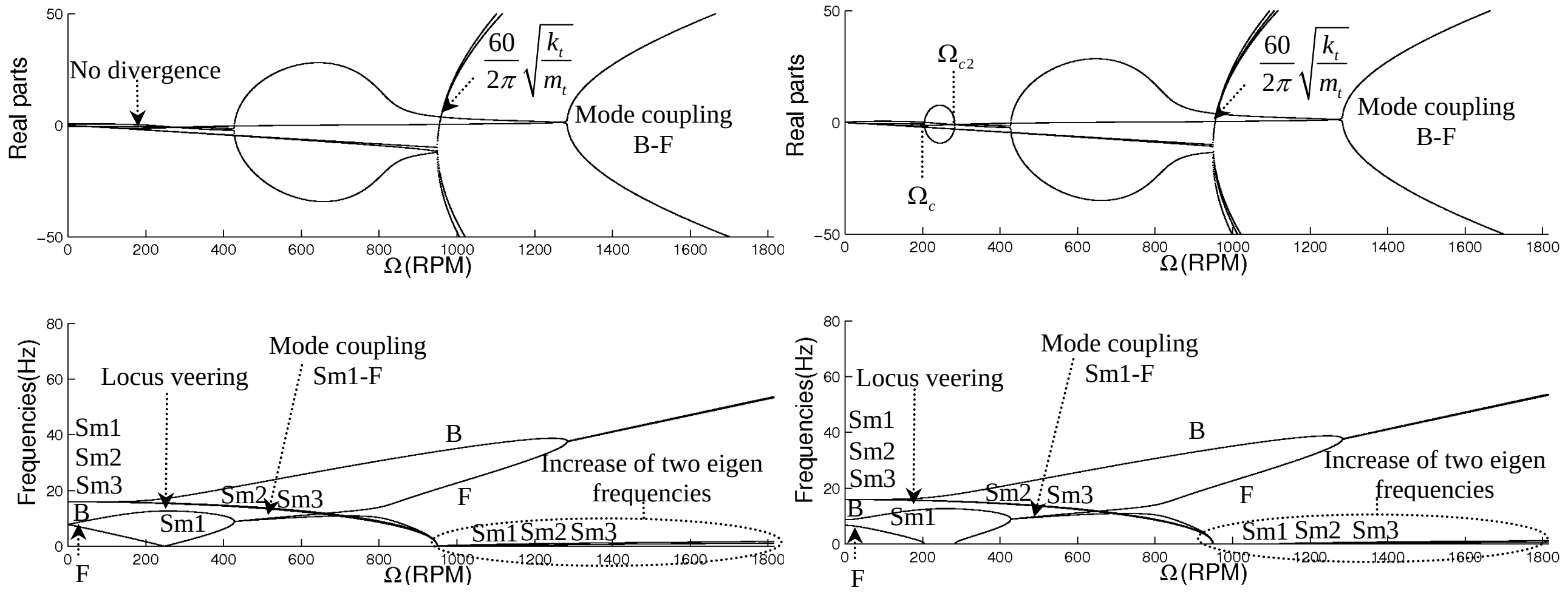

(a) (b)

Figure 13 Stability analysis of the two nodal diameter mode shape of the ring rubbed by three spring-masses having $m = 100kg$ , $k_r = k_t = 1.10^6\, N.m^{-1}$ and $\mu = 0.1$, a) separated from 60° from each other b) two being at 60° from each other and the third one at 180° from one of the latter two

Sm = Spring-mass, F = Forward, B = Backward

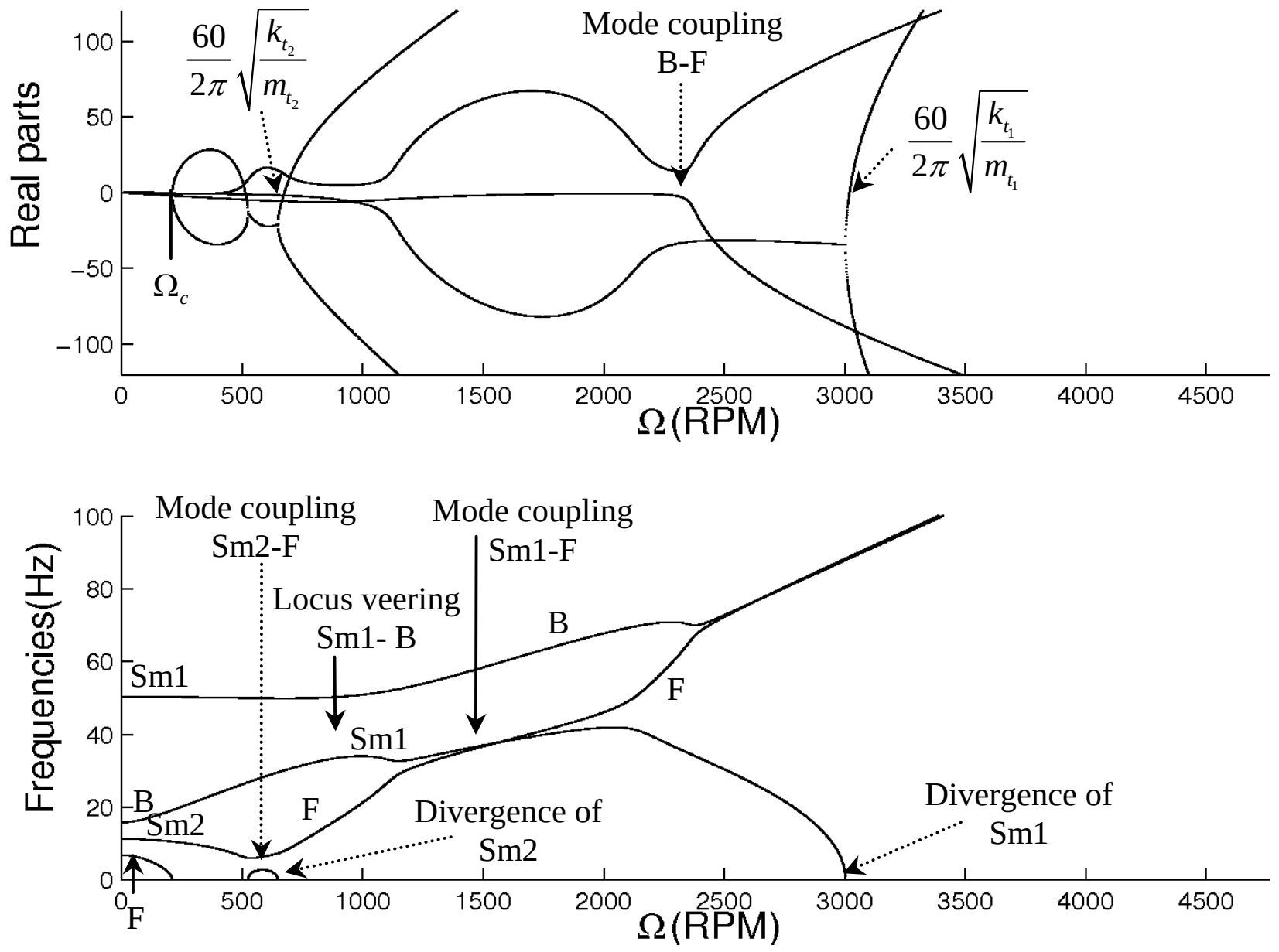

Figure 14 Stability analysis of the two nodal diameter mode shape of the ring rubbed by two spring-masses separated from 60° from each other having $m_1 = 100kg$ , $k_{r_1} = 1.10^6\,N.m^{-1}$, $k_{t_1} = 1.10^7\,N.m^{-1}$, $m_2 = 200kg$ , $k_{r_2} = 1.10^7\,N.m^{-1}$, $k_{t_2} = 1.10^6\,N.m^{-1}$ and $\mu = 0.1$

Sm = Spring-mass, F = Forward, B = Backward

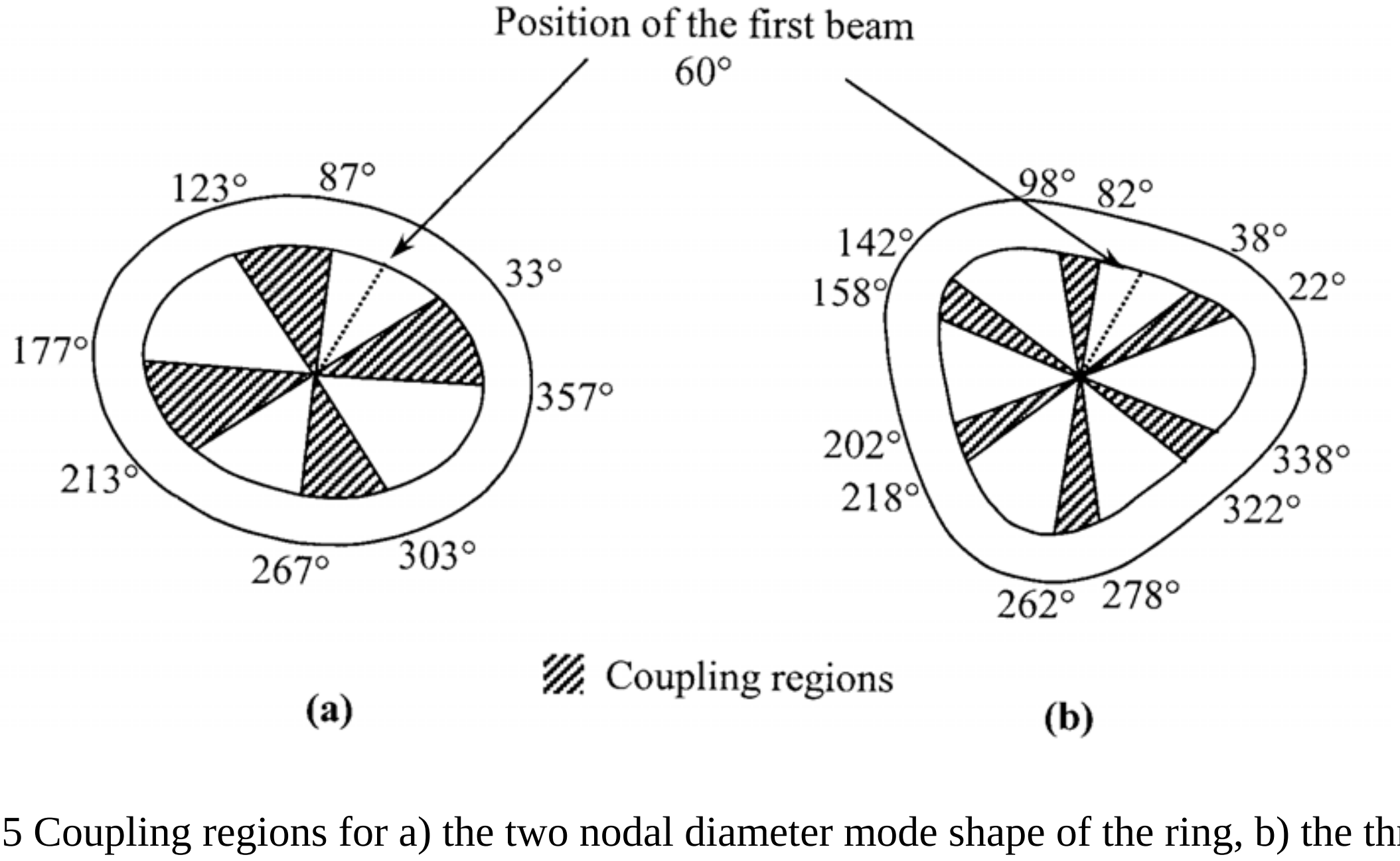

Figure 15 Coupling regions for a) the two nodal diameter mode shape of the ring, b) the three nodal diameter mode shape of the ring

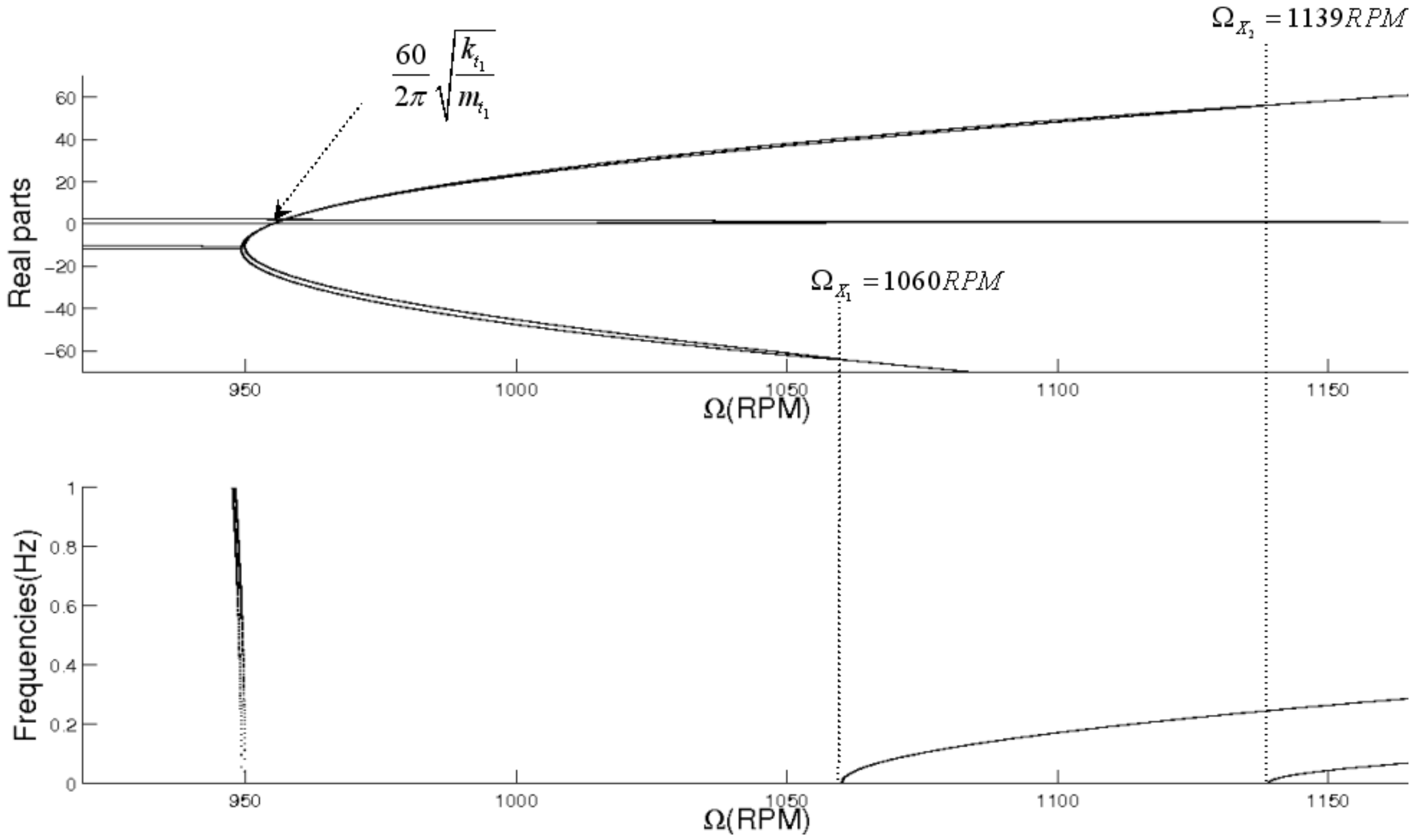

Figure 16 Coupling between two spring-masses rubbing ( $\mu = 0.1$ ) on the two nodal diameter mode shape of the ring, one being at 60° in the rotating frame and the other, at 120°, with $m_1 = 100kg$ and $k_r = k_t = 1.10^6 N.m^{-1}$

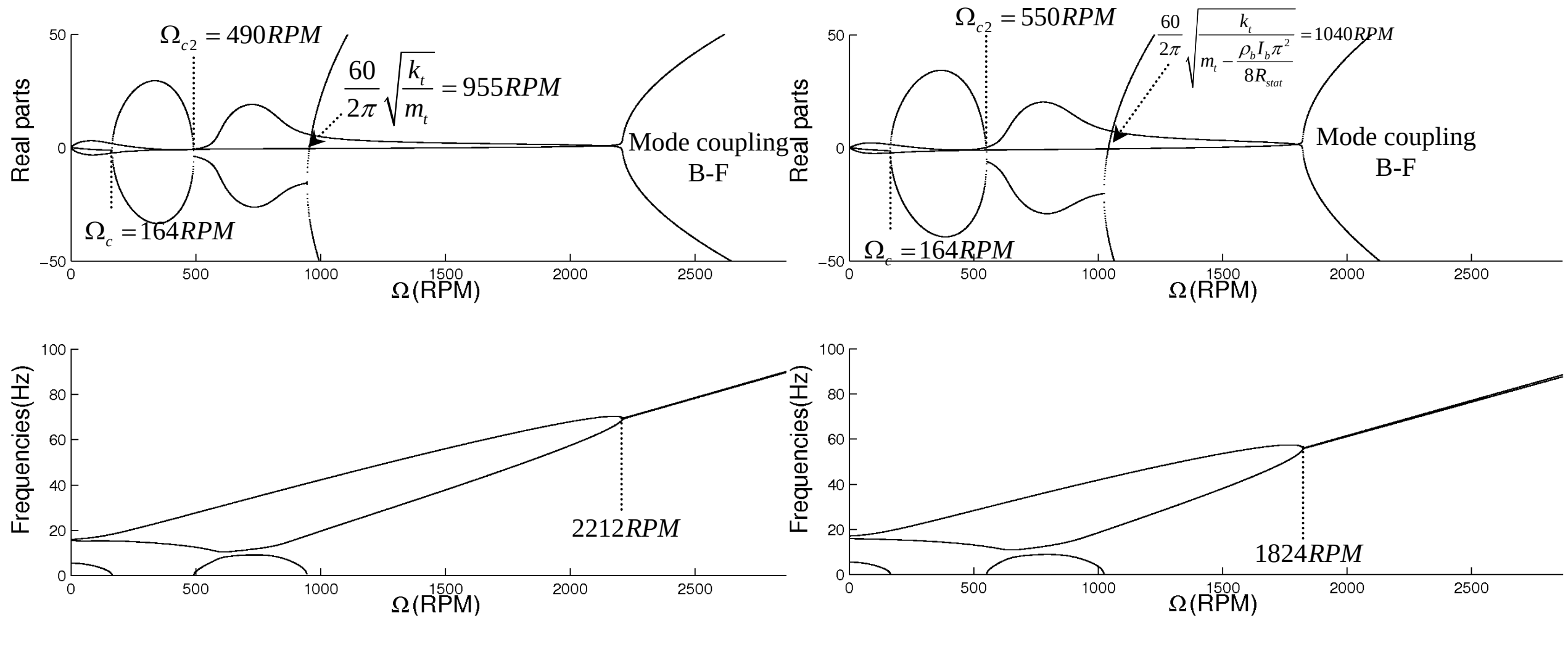

Figure 17 Stability analysis for the two nodal diameter mode shape of the ring rubbed by a) one spring-mass having two degrees of freedom b) one beam having two degrees of freedom

Sm = Spring-mass, bt= flexure motion of the beam F = Forward, B = Backward

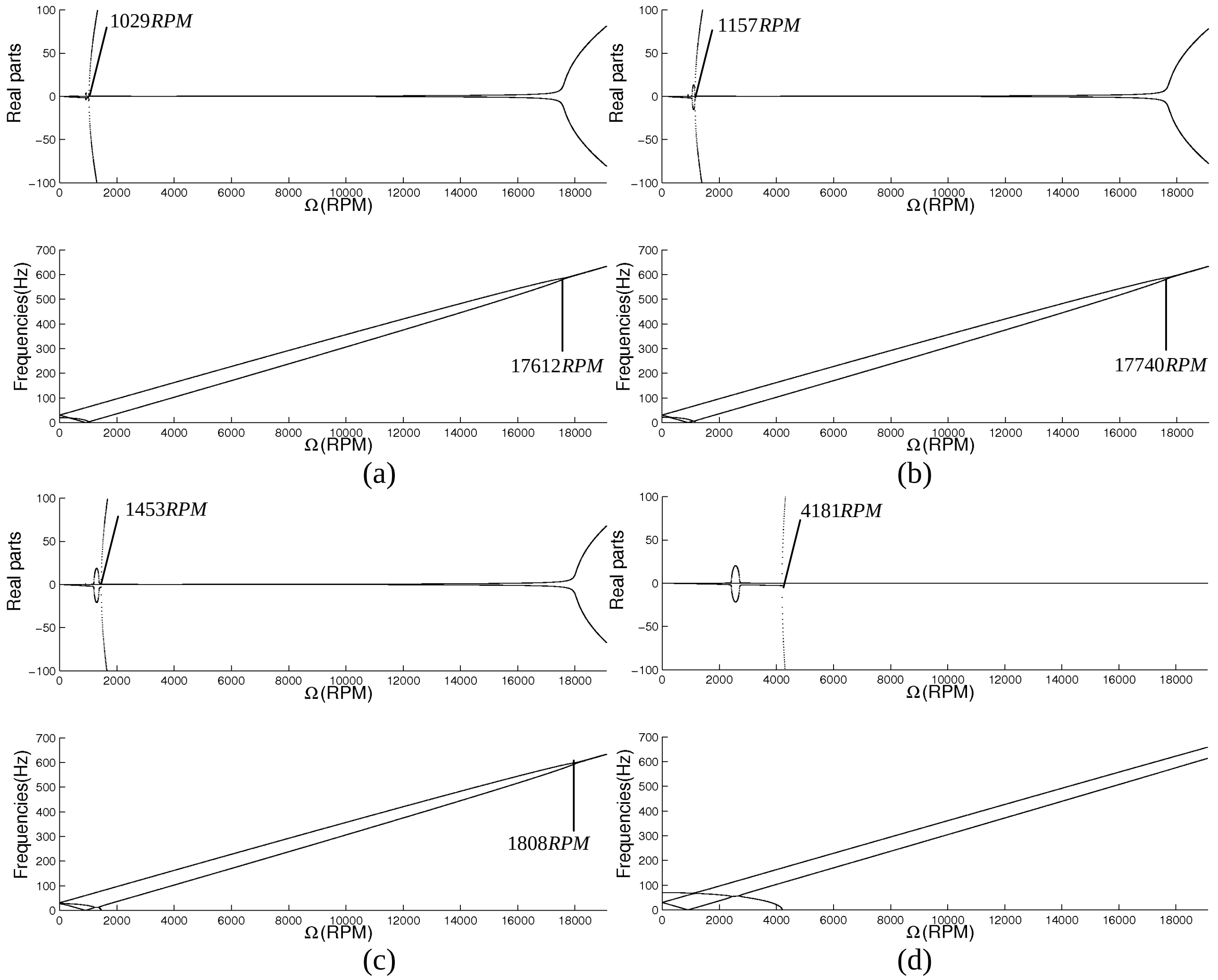

Figure 18 Stability analysis of the two nodal diameter mode shape of the ring (30Hz) rubbed by one beam (20Hz) at a) $\alpha = 0°$, b) $\alpha = 5°$, c) $\alpha = 10°$ and d) $\alpha = 89°$

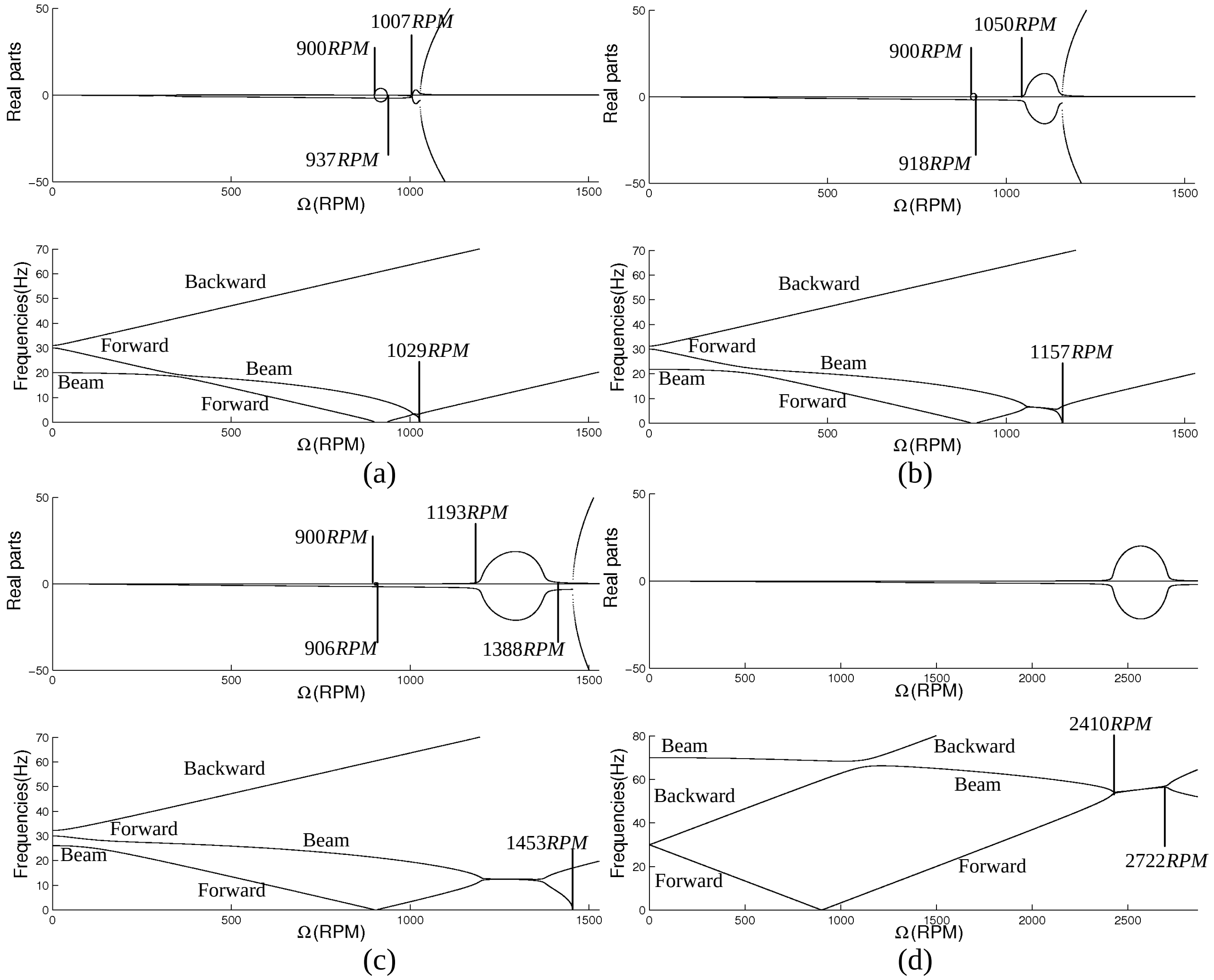

Figure 19 Zooms associated with Fig. 18 – evolution of the forward mode shape divergence of the ring and mode couplings as a function of $\alpha$ , for a) $\alpha = 0°$, b) $\alpha = 5°$, c) $\alpha = 10°$ and d) $\alpha = 89°$

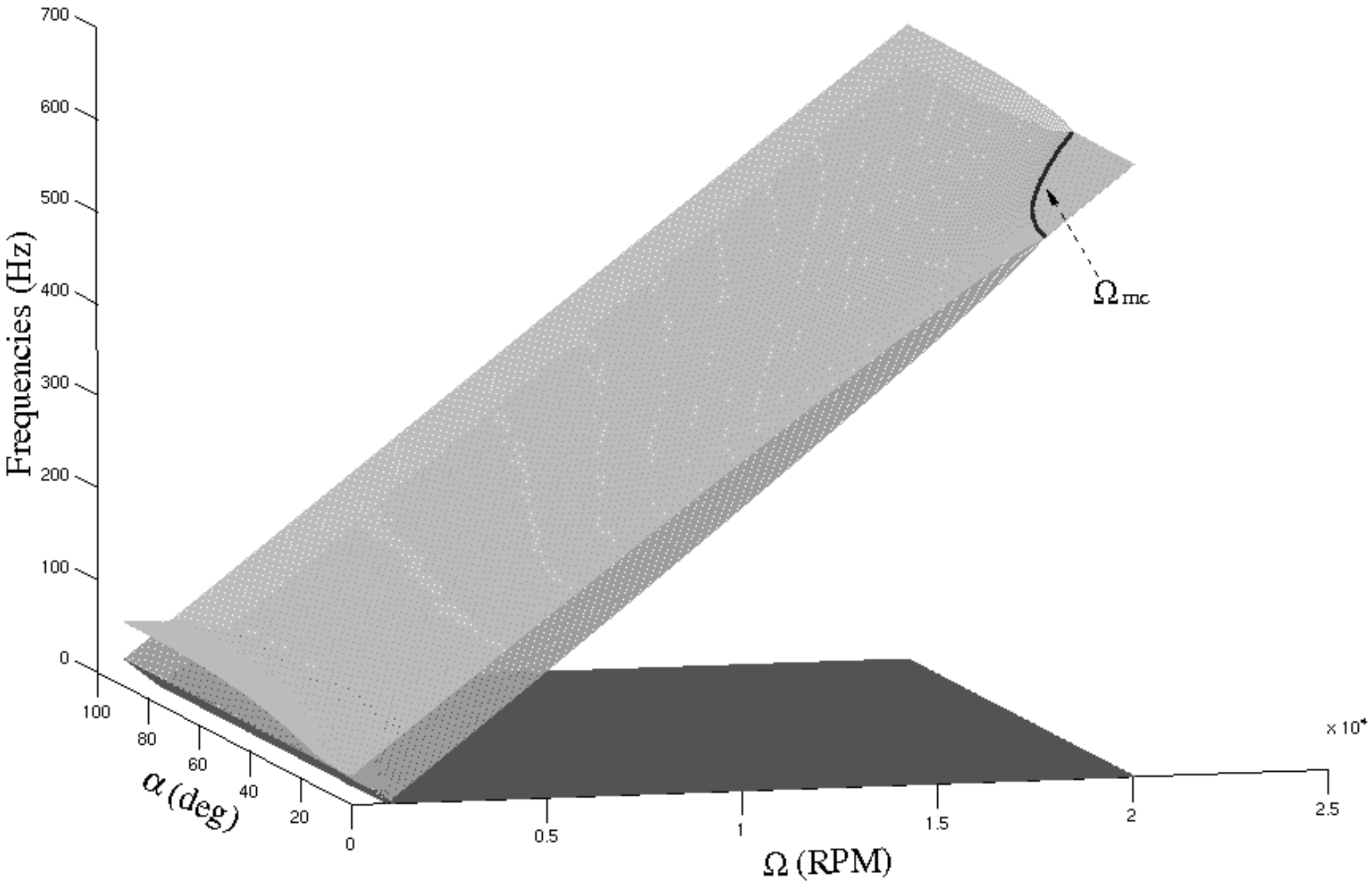

Figure 20 Campbell diagram of the two nodal diameter mode shape of the ring (30Hz) rubbed by one beam (20Hz), as a function of $\alpha$ - evolution of $\Omega_{mc}$

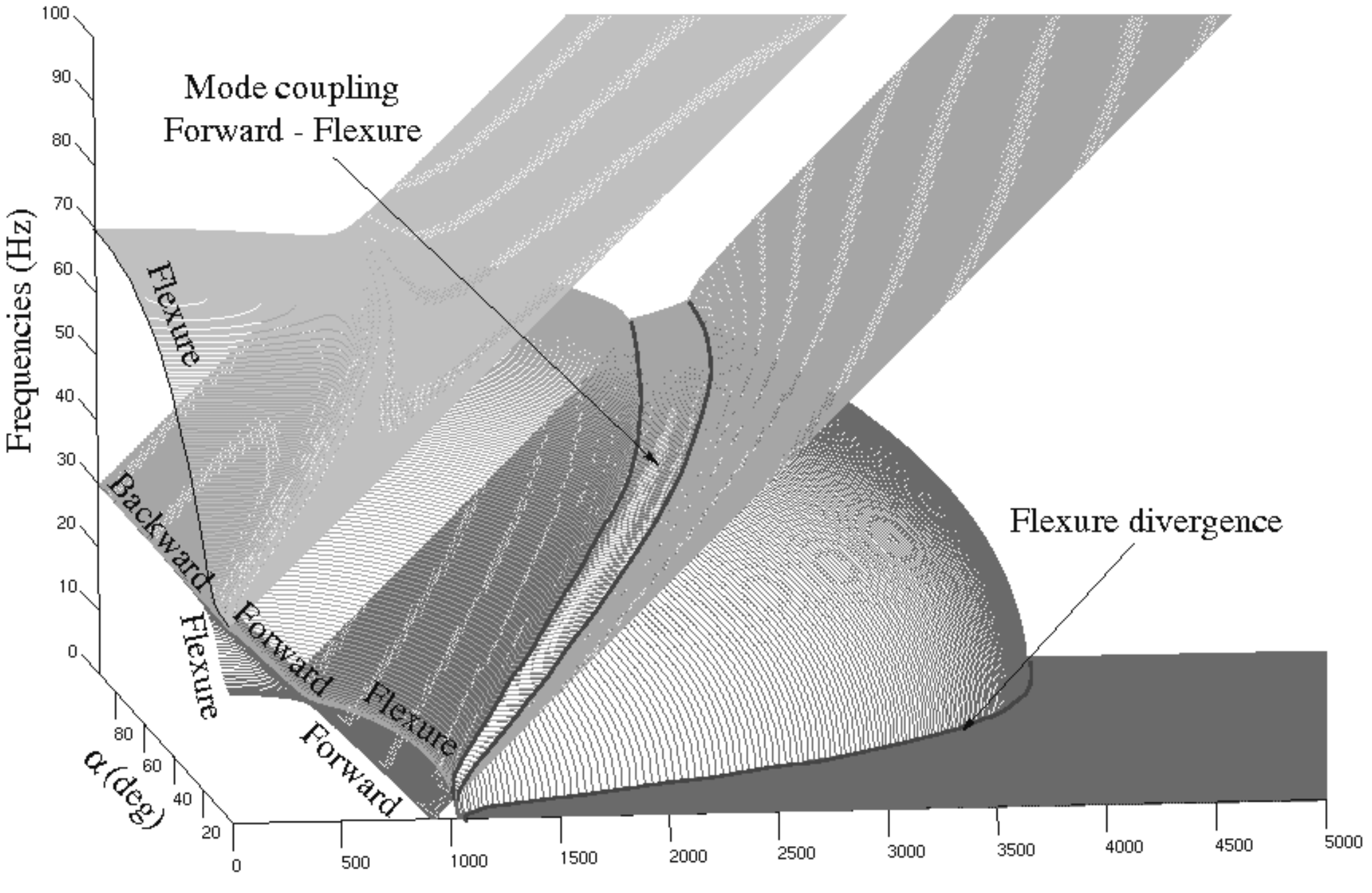

Figure 21 Campbell diagram of the two nodal diameter mode shape of the ring (30Hz) rubbed by one beam (20Hz), as a function of $\alpha$ - position of the mode couplings between the forward mode shape and the flexure motion of the beam as well as the evolution of the divergence of this latter degree of freedom